\documentclass[acmsmall,natbib=false]{acmart}
\pdfoutput=1

\AtBeginDocument{%
	}

\setcopyright{acmcopyright}
\copyrightyear{2023}
\acmYear{2023}
\acmDOI{XXXXXXX.XXXXXXX}

\acmJournal{TECS}
\acmVolume{42}
\acmNumber{42}
\acmArticle{42}
\acmMonth{8}

\usepackage{xcolor}
\usepackage{tikz, pgfplots}
\usepgfplotslibrary{statistics}
\pgfplotsset{compat=1.18}
\usetikzlibrary{matrix, positioning, fit, colorbrewer, decorations.pathmorphing}
\usepgfplotslibrary{colorbrewer}
\usetikzlibrary{circuits.logic}
\usetikzlibrary{circuits.logic.US}
\usetikzlibrary{shapes.multipart}
\pgfdeclarelayer{bg}    
\pgfsetlayers{bg,main}  
\usetikzlibrary{shapes.symbols}
\usetikzlibrary{shadows}
\usetikzlibrary{shapes.misc}
\usetikzlibrary{shapes}

\newcommand{\defineColorSet}[3][C]{%
	\colorlet{#2Color}{#3-#1}	
	\colorlet{#2Dark}{#3-M}	
	\colorlet{#2Line}{#3-K}	
	\colorlet{#2Text}{#3-L}	
}

\defineColorSet{MembranePotentialSlope}{Blues}
\defineColorSet{AEQ}{Blues}
\defineColorSet{Generic}{Greys}
\defineColorSet{MembranePotential}{Greens}
\defineColorSet{ROM}{Oranges}
\defineColorSet[E]{ClassificationU}{Purples}

\tikzset{
	spike/.style={red, ultra thick},
	box/.style={draw, rectangle, align=center, line width=1.1pt},
	circ/.style={draw, circle, fill=white},
	rom/.style={box, fill=ROMColor, draw=ROMDark},
	aeq/.style={box, fill=AEQColor, draw=AEQDark},
	memPot/.style={box, fill=MembranePotentialColor, draw=MembranePotentialDark},
	genBox/.style={box, fill=GenericColor, draw=GenericDark},
	classU/.style={box, fill=ClassificationUColor, draw=ClassificationUDark},
}

\tikzset{mymatrix/.style={matrix of nodes,nodes in empty cells,left,row sep=-\pgflinewidth,column sep=-\pgflinewidth,text depth=0.5ex,text height=2ex,
		nodes={align=center,text width=20pt,minimum height=20pt,anchor=center,inner sep=0pt,outer sep=0pt, rectangle,text=MembranePotentialSlopeText, thick, draw=MembranePotentialSlopeLine, fill=MembranePotentialSlopeColor}}}
	
	\tikzset{%
		fancyMatrix/.style={%
			left,
			row sep=-\pgflinewidth,
			column sep=-\pgflinewidth,
			text depth=0.5ex,
			text height=1.5ex
		},
		matrixNode/.style={%
			align=center,
			text width=25pt,
			minimum height=25pt,
			anchor=center,
			inner sep=0pt,
			outer sep=0pt, 
			rectangle,
			thick, 
		},
		memPotMatrixNode/.style={%
			matrixNode,
			draw=MembranePotentialLine, 
			fill=MembranePotentialColor
		},
		imemNode/.style={
			memPotMatrixNode,
			text width=20pt,
			minimum height=20pt
		},
		queue/.style={%
			matrix of nodes, 
			nodes in empty cells, 
			outer sep=0pt
		},
		imem/.style={
			queue,
			fancyMatrix, 
			nodes=imemNode
		},
		aeqMatrixNode/.style={%
			matrixNode,
			draw=AEQLine,
			fill=AEQColor
		},
		aeqMemNode/.style={%
			aeqMatrixNode,
			text width=22pt,
			minimum height=15pt		
		},
		aeqMem/.style={
			queue,
			fancyMatrix,
			nodes=aeqMemNode
		}
	}

\usepackage[utf8]{inputenc}

\usepackage[backend=biber, hyperref, style=numeric, maxnames=70, minnames=4, giveninits=true, maxcitenames=2, mincitenames=1, bibencoding=utf8, isbn=true, sorting=none]{biblatex}

\hypersetup{
	pdfauthor={Patrick Plagwitz, Frank Hannig, Jürgen Teich, Oliver Keszocze},
	pdftitle={To Spike or Not to Spike? A Quantitative Comparison of SNN and CNN FPGA Implementations}
}

\usepackage{ifthen}

\usepackage{siunitx}
\usepackage{xspace}
\usepackage{scalefnt}

\usepackage{pgfplotstable}
\usepackage{booktabs}
\usepackage{array}
\usepackage{colortbl}
\pgfplotstableset{
	every head row/.style={before row=\toprule,after row=\midrule},
	every last row/.style={after row=\bottomrule},
	assign column name/.style={/pgfplots/table/column name={\textbf{#1}}},
}

\usepackage{xcolor}
\usepackage{tikz, pgfplots}
\usepgfplotslibrary{statistics}
\pgfplotsset{compat=1.18}
\usetikzlibrary{matrix, positioning, fit, colorbrewer}
\usetikzlibrary{circuits.logic}
\usetikzlibrary{circuits.logic.US}
\pgfdeclarelayer{bg}    
\pgfsetlayers{bg,main}  

\usepackage{setspace}
\usepackage{csquotes}

\usepackage{acronym}
\newacro{AE}{Address Event}
\newacro{AEQ}{Address Event Queue}
\newacro{SNN}{Spiking Neural Network}
\newacro{ADL}{Architecture Description Language}
\newacro{CGRA}{Coarse-Grained Reconfigurable Array}
\newacro{HBM}{High-Bandwidth Memory}
\newacro{VTA}{Versatile Tensor Accelerator}
\newacro{GEMM}{General Matrix-Multiply}
\newacro{BERT}{Bidirectional Encoder Representations from Transformers}
\newacro{NLP}{Natural Language Processing}
\newacro{GELU}{Gaussian Error Linear Unit}
\newacro{QAT}{Quantization-Aware Training}
\newacro{PE}{Processing Element}
\newacro{LA}{Linear Approximator}
\newacro{TTFS}{Time-To-First Spike}

\newacro{MAC}{Multiply Accumulate}
\newacro{ML}{Machine Learning}
\newacro{DPU}{Deep-learning Processing Unit}
\newacro{ANN}{Artificial Neural Network}
\newacro{NoC}{Network on Chip}
\newacro{STDP}{Spike-Timing-Dependent Plasticity}
\newacro{BNN}{Binarized Neural Network}
\newacro{DL}{Deep Learning}
\newacro{DNN}{Deep Neural Network}
\newacro{CNN}{Convolutional Neural Network}
\newacro{LSTM}{Long Short-Term Memory}
\newacroplural{LSTM}{Long Short-Term Memories}
\newacro{GRU}{Gated Recurrent Unit}
\newacro{RNN}{Recurrent Neural Network}
\newacro{NN}{Neural Network}
\newacro{AES}{Advanced Encryption Standard}
\newacro{ALU}{Arithmetic Logic Unit}
\newacro{API}{Application Programming Interface}
\newacro{AST}{Abstract Syntax Tree}
\newacro{ASIC}{Application-Specific Integrated Circuit}
\newacro{ASIP}{Application-Specific Instruction Set Processor}
\newacro{BDG}{Boolean Data-Flow Graph}
\newacro{BRAM}{Block RAM}
\newacro{CSC}{Color Space Conversion}
\newacro{DFG}{Data-Flow Graph}
\newacro{DMA}{Direct Memory Access}
\newacro{DPU}{Deep Learning Processing Unit}
\newacro{DSE}{Design Space Exploration}
\newacro{DSL}{Domain-Specific Language}
\newacro{DSP}{Digital Signal Processor}
\newacro{DUT}{Device Under Test}
\newacro{ECC}{Error-Correcting Code}
\newacro{EFA}{Extended Finite Automata}
\newacro{ESL}{Electronic System Level}
\newacro{FF}{Flip-Flop}
\newacro{FFT}{Fast-Fourier Transform}
\newacro{FIFO}{First Input First Output}
\newacro{FOV}{Field of View}
\newacro{FPGA}{Field-Programmable Gate Array}
\newacro{FSM}{Finite State Machine}
\newacro{FPU}{Floating-Point Unit}
\newacro{fps}{frames per second}
\newacro{GPP}{General-Purpose Processor}
\newacro{GPU}{Graphics Processing Unit}
\newacro{TPU}{Tensor Processing Unit}
\newacro{IoT}{Internet of Things}
\newacro{ONNX}{Open Neural Network Exchange}
\newacro{GPIO}{General-Purpose Input/Output}
\newacro{HDL}{Hardware Description Language}
\newacro{HIPAcc}[\protect\hipacc{}]{Heterogeneous Image Processing Acceleration}
\newacro{HLL}{High-Level Language}
\newacro{HLS}{High-Level Synthesis}
\newacro{HPF}{High-Pass Filter}
\newacro{IDCT}{Inverse Discrete Cosine Transform}
\newacro{IIR}{Infinite impulse response}
\newacro{ILA}{Integrated Logic Analyzer}
\newacro{IPB}{Intellectual Property Block}
\newacro{ISA}{Instruction Set Architecture}
\newacro{LUT}{Lookup table}
\newacro{MB}{MicroBlaze}
\newacro{MOEA}{Multiobjective Evolutionary Algorithm}
\newacro{OpenCV}{Open Source Computer Vision}
\newacro{PSoC}{Programmable System-on-Chip}
\newacro{SoC}{System-on-Chip}
\newacro{PL}{Programmable Logic}
\newacro{PPM}{Portable Pixmap}
\newacro{PS}{Processing System}
\newacro{RTL}{Register-Transfer Level}
\newacro{RLD}{Run-length Decoding}
\newacro{SDF}{Synchronous Data-Flow}
\newacro{SDL}{System Description Language}
\newacro{SFU}{Special Function Unit}
\newacro{SoC}{System-on-Chip}
\newacro{TLC}{Target Language Compiler}
\newacro{UML}{Unified Modeling Language}

\newacro{OS}{Operating System}
\newacro{MMU}{Memory Management Unit}
\newacro{RISC}{Reduced Instruction Set Computing}
\newacro{DAG}{Directed Acylic Graph}
\newacro{IR}{Intermediate Representation}
\newacro{TICG}{Target-Independent Code Generator}
\newacro{DSA}{Direct Stream Access}
\newacro{ASA}{Asynchronous Stream Access}
\newacro{IDCT}{Inverse Discrete Cosine Transform}
\newacro{CSC}{Colorspace Conversion}

\usepackage{caption}
\usepackage{subcaption}
\newcommand{\tblheader}[1]{{\bfseries #1}}
\newcommand{\snnname}[2]{$\textsc{SNN#2}_\textsc{#1}$}
\newcommand{\snneightpacking}{\snnname{COMPR.}{8}}
\newcommand{\snnfourpacking}{\snnname{COMPR.}{4}}
\newcommand{\snnfoursvhn}{\snnname{SVHN}{4}}
\newcommand{\snntwosvhn}{\snnname{SVHN}{2}}
\newcommand{\snneightsvhn}{\snnname{SVHN}{8}}
\newcommand{\snnsixteensvhn}{\snnname{SVHN}{16}}
\newcommand{\snnfourcifar}{\snnname{CIFAR}{4}}
\newcommand{\snntwocifar}{\snnname{CIFAR}{2}}
\newcommand{\snneightcifar}{\snnname{CIFAR}{8}}
\newcommand{\snnsixteencifar}{\snnname{CIFAR}{16}}
\newcommand{\snneightbram}{\snnname{BRAM}{8}}
\newcommand{\snnfourbram}{\snnname{BRAM}{4}}
\newcommand{\snneightlutram}{\snnname{LUTRAM}{8}}
\newcommand{\snnfourlutram}{\snnname{LUTRAM}{4}}

\newcommand{\snnsixteenpacking}{\snnname{COMPR.}{16}}

\newcommand{\snnonebram}{\snnname{BRAM}{1}}
\newcommand{\cnnname}[1]{$\textsc{CNN}_{#1}$}

\newcommand{\zcu}{ZCU102}

\newcommand{\ie}{\mbox{i.e.,}\xspace}
\newcommand{\eg}{\mbox{e.g.,}\xspace}

\newcolumntype{N}{S[table-format=5.0, group-separator={,}]}

\begin{document}

\title{To Spike or Not to Spike? A Quantitative Comparison of SNN and CNN FPGA Implementations}


\author{Patrick Plagwitz}
\email{patrick.plagwitz@fau.de}
\author{Frank Hannig}
\email{frank.hannig@fau.de}
\author{Jürgen Teich}
\email{juergen.teich@fau.de}
\author{Oliver Keszocze}
\email{oliver.keszoecze@fau.de}
\affiliation{%
	\institution{\newline Department of Computer Science, Friedrich-Alexander-Universität Erlangen-Nürnberg (FAU)}
	\streetaddress{Cauerstr.~11}
	\city{91058~Erlangen}
	\country{Germany}
}


\begin{abstract}
Convolutional Neural Networks (CNNs) are widely employed to solve various problems, e.g., image classification.
Due to their compute- and data-intensive nature, CNN accelerators have been developed as ASICs or on FPGAs.
Increasing complexity of applications has caused resource costs and energy requirements of these accelerators to grow.
Spiking Neural Networks (SNNs) are an emerging alternative to CNN implementations, promising higher resource and energy efficiency.
The main research question addressed in this paper is whether SNN accelerators truly meet these expectations of reduced energy requirements compared to their CNN equivalents.
For this purpose, we analyze multiple SNN hardware accelerators for FPGAs regarding performance and energy efficiency.
We present a novel encoding scheme of spike event queues and a novel memory organization technique to improve SNN energy efficiency further.
Both techniques have been integrated into a state-of-the-art SNN architecture and evaluated for MNIST, SVHN, and CIFAR-10 datasets and corresponding network architectures on two differently sized modern FPGA platforms.
For small-scale benchmarks such as MNIST, SNN designs provide rather no or little latency and energy efficiency advantages over corresponding CNN implementations.
For more complex benchmarks such as SVHN and CIFAR-10, the trend reverses.
\end{abstract}

\ccsdesc[500]{Hardware~Hardware accelerators}
\ccsdesc[500]{Computer systems organization~Neural networks}
\ccsdesc[500]{Hardware~Power estimation and optimization}

\keywords{Spiking Neural Networks, FPGA, Convolutional Neural Networks}

\received{20 February 2007}
\received[revised]{12 March 2009}
\received[accepted]{5 June 2009}
\maketitle

\section{Introduction}
\acp{SNN} and traditional \acp{ANN} represent two diverging research directions.
While, \eg \acp{CNN} and transformer-based networks have come a long way from the initial idea of implementing a biological brain as an algorithm, \ac{SNN} research still strives to find architectures based on biologically plausible neuron models \cite{snnsurvey,stochastic}.
They are defined by their non-temporal representation of neurons as weight matrices.
We use the term \enquote{traditional \acp{ANN}} when referring to neural network counterparts of \acp{SNN} that are non-spiking.
Traditional \acp{ANN} are easily justified by their vast successes in various computing domains like image or audio processing or natural language when it comes to generative models.
Advances in this field have also been driven by ever more powerful hardware for parallelized matrix-matrix multiplications, which is the central compute-intensive component of both learning and inference in these types of networks.
Platforms, including \acp{GPU} but also \acp{FPGA}, have proven to be viable targets for specialized \ac{ANN} accelerators.

On the other hand, \acp{SNN} feature properties that make them particularly interesting for hardware acceleration.
For example, they are inherently event-driven, rendering them suitable for applications where sensor data is generated in an event-driven manner \cite{stochastic}.
Another consequence of this fact is that only spiked neurons need to be considered for computation, \ie a network input that generates only a few or even no spikes can be evaluated in a very short amount of time.
This is in contrast to \acp{ANN}, where all neurons must be computed unless optimization techniques like pruning are employed.
In \acp{SNN}, pruning can be achieved implicitly.
Moreover, occurring spikes can be evaluated multiplier-less, providing even more potential for cost and energy savings.

Finally, the biological focus can also be reduced, creating a hybrid approach of hardware-friendly \acp{SNN}.
For example, the neuron model does not need to be completely biologically accurate to produce a good network quality for classification tasks.
The popular integrate-and-fire model \cite{bouvier} allows spikes to be represented as bits and all neuron computations to be executed completely multiplier-free, a property explained later in more detail.

An important question to be tackled is whether accelerators for hardware-friendly \acp{SNN} truly outperform traditional \ac{ANN} accelerators in terms of performance, resource cost, and energy requirements.
In this work, we focus on \acp{CNN} and \acp{SNN} that include convolutional layers and investigate the subject by employing image classification as the use case.
To make comparisons fair, various FPGA-specific and NN-related metrics must be taken into account.
These include FPGA resource usage, classification accuracy, target platform, and network architecture.
In the following, we summarize our main contributions.

\subsection*{Main Contributions:}

\begin{itemize}
\item Comparing both SNN and CNN state-of-the-art accelerator designs to answer the question if \ac{FPGA}-based \acp{SNN} provide significant performance improvements over \ac{CNN} counterparts.\\[-.75em]
\item Providing an analysis and comparing energy efficiency for FPGA-based SNN and CNN implementations on on differently sized AMD Xilinx devices.\\[-.75em]
\item Proposition of a novel spike queue memory architecture and spike encoding for a state-of-the-art SNN approach in order to reduce its memory footprint and energy requirements.
\end{itemize}

\noindent

The remainder of this paper is structured as follows:
Section~\ref{sec:background} provides the required fundamentals of \acp{SNN} and \acp{CNN} and discusses the related work in these fields.
Section~\ref{sec:archs} then gives an in-depth introduction to the two \ac{NN} architectures/frameworks used for our research: the state-of-the-art \ac{SNN} accelerator by~\citeauthor{sommer}~\cite{sommer}, which serves as a basis for our novel architectural contributions later in Section~\ref{sec:improvements}, and the FINN \ac{CNN} framework~\cite{Blott2018}.
Experimental results are presented in Section~\ref{sec:experiments}.
Here, directly comparing the energy efficiency of the \ac{SNN} accelerator against FINN-based \acp{CNN} on the MNIST benchmark on an PYNQ-Z1 board reveals that \acp{SNN} are no simple drop-in replacement for \acp{CNN}.
Consequently, this section conducts various experiments to identify where and how optimizations might alleviate this issue.
In Section~\ref{sec:improvements}, two optimizations (a novel spike encoding and spike memory architecture) are presented and evaluated.
As additional benchmarks, the SVHN and CIFAR-10 data sets are used.
To evaluate the scalability of our \ac{SNN} accelerator approach, a larger FPGA board (ZCU102) is used as a second target device.
Finally, Section~\ref{sec:conclusion} concludes the paper.


\section{Background}\label{sec:background}


In this section, fundamentals regarding \acp{SNN} are given.
Furthermore, the section highlights the differences between \acp{SNN} and CNNs regarding hardware acceleration.

\subsection{Spiking Neural Networks}
Contrary to conventional \acp{NN} models, \aclp{SNN} are biologically motivated. 
As a result, they encode activations not with real-valued numbers but with sequences of binary spikes.
An extensive set of \ac{SNN} models has emerged to tradeoff between biological plausibility and model complexity. On the one hand, the Hodgkin-Huxley model describes the complex electrochemical processes of biological neurons, but is prohibitively expensive in its computational cost \cite{izhikevich}.
On the other hand, the \textit{Integrate-and-Fire (IF)} model only loosely models biological neurons, but is better suited for hardware implementations~\cite{bouvier}.

Here, two conflicting schools of thought can be identified: neurological research and the application of \acp{SNN} in a practical context.
In general, a clear assignment of \ac{SNN} implementations to one of these is impossible.
Many implementations add biologically inspired computational primitives to the standard IF model, hoping that more \emph{bioplausibility} improves performance (in terms of classification accuracy or computational performance).

Apart from the neuron model, the \textit{spike encoding} (\ie how spikes encode numbers) is a defining characteristic of SNN accelerators.
Objectives affected by it are training time, accuracy, and classification latency because some encodings allow for similar accuracy to be achieved in fewer so-called \textit{algorithmic time steps} but are more complicated to implement.

\subsubsection{The Integrate-and-Fire (IF) Model}
When comparing the IF model to how neurons are modeled in standard non-spiking \acp{NN}, two major differences can be identified:
\begin{itemize}
\item In the IF model, neuron activations are represented by binary spikes.
In standard \acp{NN}, neuron activations are real-valued: a large activation is represented by a large numeric value and vice versa.
\item Furthermore, in the IF model, neurons have an internal state called the \textit{membrane potential} \(V_m\).
Their activation is dependent on their membrane potential.
In standard \acp{NN}, neurons are not stateful, and their output only depends directly on their inputs.
\end{itemize}

The \textit{IF model} depends on neurons being evaluated repeatedly in discrete \textit{algorithmic time steps} \(t\).
If a spike arrives at neuron $j$ via its synapse $i$ (having weight $w_i$) at time step $t$, the weight of the synapse is added to the neuron's membrane potential, \ie $V_{m_j}(t)$ is computed as

\begin{equation}
V^l_{m_j}(t) = \begin{cases}
  0 & \text{ if } V^l_{m_j}(t-1) > V_t \\
  V^l_{m_j}(t-1) + \sum_i{w_{i,j} \cdot x_i^{l-1}(t-1)} & \text{ otherwise}
 \end{cases}.
\label{eq:ifvm}
\end{equation}

Here, \(l\) is the layer of the currently evaluated neuron and \(l-1\) the previous layer where the spike originated.
Also, \(V_t\) denotes a threshold value for the membrane potential.
Whenever it is crossed, the neuron will (a) reset its membrane potential \(V_{m_j}\) to 0, and (b) generate a spike itself, which then travels to all connected neurons.
The neuron output \(x\) defines thus whether a spike has occurred at time step \(t\):

\begin{equation}
x^l_j(t) = \begin{cases}
  1 & \text{ if } V^l_{m_j}(t) > V_t \\
  0 & \text{ otherwise.}
 \end{cases}
\label{eq:ifx}
\end{equation}

A biologically more accurate but also more compute-intensive extension of the IF model is the \emph{leaky IF model}, where a constant leakage term \(\lambda\) is introduced in Eq.~(\ref{eq:ifvm}).
Here, the membrane potential constantly decreases as a function of \(\lambda\) \cite{bouvier}.
However, this paper considers only the IF model without leakage due to its hardware-friendliness.

To implement these equations in hardware, operators and intermediate memories must be considered.
First, the memory potentials must be stored.
\acp{SNN} are inherently temporal: they have an internal state and their output is dependent on all previous time steps instead of the current input (e.g., image to classify) only.
Also, multiplications do not need to be carried out at all: as the variables $x_i^{l}$ only take the values $0$ and $1$, they effectively serve as selector variables indicating that the weight should be added ($x_i^l=1$) or not ($x_i^l=0$).
This is a significant inherent difference between \acp{SNN} and \acp{CNN}.
Instead of multiplying all activations all the time, only additions need to be performed whenever the \textit{sparse} feature maps contain spikes \cite{mostafa}.
The question of whether the tradeoff between memory requirements and decreased computational cost leads to more efficient hardware designs shall be answered in this paper.

\subsubsection{Spike Encodings}
Another essential characteristic of an \ac{SNN} implementation is the way spikes are encoded.
Biologically, the significance of a spike is determined by the time it appears in connection with preceding and subsequent spikes, thereby affecting the membrane potential of a neuron.
Several encoding methods have been proposed to try to capture this principle \cite{neuralcoding}.
Two commonly used ones in hardware accelerators are: Direct temporal coding, rate coding \cite{neuralcoding}, and \ac{TTFS} coding~\cite{rueckauer2}.

Rate coding requires neurons to estimate the \textit{firing rate} of connected neurons by averaging spikes over a time window.
The size of this window has significance for the hardware resources and execution time needed to arrive at a stable value for the firing rate until feed-forward computations can be performed.
Likewise, a larger time window allows for higher \ac{SNN} accuracy after training.
Therefore, a tradeoff exists between timing error robustness, latency, and accuracy when choosing the time window \cite{neuralcoding}.

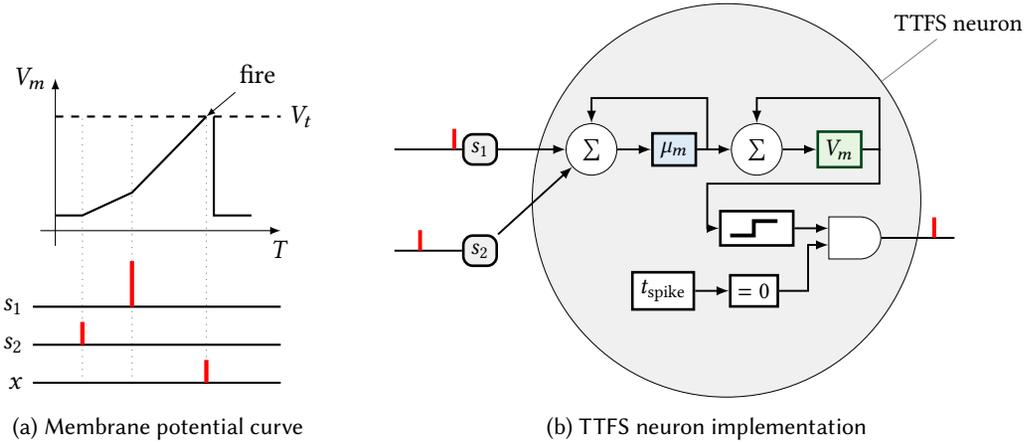
\begin{figure}
	\centering
	\begin{subfigure}{.35\linewidth}
		\centering
		\begin{tikzpicture}
			\draw[-latex] (-.2,0) -- node[below,pos=1] {$T$} (3,0);
			\draw[-latex] (0,-.2) -- node[left,pos=1] {$V_m$} (0,2);
			\draw[thick] (-.3,-1) -- node[left, pos=0] {$s_1$} (3,-1);
			\draw[thick] (-.3,-1.5) -- node[left, pos=0] {$s_2$}(3,-1.5);
			\draw[thick] (-.3,-2) -- node[left, pos=0] {$x$} (3,-2);
			
			\coordinate (s1spikepos) at ($ (-.3,-1)!.4!(3,-1)$);
			\coordinate (s2spikepos) at ($ (-.3,-1.5)!.2!(3,-1.5)$);
			\coordinate (xspikepos) at ($ (-.3,-2)!.7!(3,-2)$);

			\draw[spike] (s1spikepos) -- ++ (0,.6);
			\draw[spike] (s2spikepos) -- ++ (0,.3);
			\draw[spike] (xspikepos) -- ++ (0,.3);
			
			\path[draw, thick] let \p1 = (s1spikepos),
			\p2 = (s2spikepos),
			\p3 = (xspikepos)
			in
			(0,.2) -- coordinate[pos=1] (bend1) (\x2,.2) -- coordinate[pos=1] (bend2) (\x1, .5) --  (\x3, 1.5) ++(.1,0) -- ++(0,-1.3) -- ++(.5,0);
			
			\draw[dashed, thick] (0,1.5) -- node[right, pos=1] {$V_t$} (3,1.5);
			
			\begin{pgfonlayer}{bg}
				\path[draw, gray, dotted] let \p1 = (s1spikepos)
				in (\x1, -2) -- (\x1, 1.5);
				\path[draw, gray, dotted] let \p1 = (s2spikepos)
				in (\x1, -2) -- (\x1, 1.5);
				\path[draw, gray, dotted] let \p1 = (xspikepos)
				in (\x1, -2) -- coordinate[above, pos=1, pin={[black, pin edge={solid, black, latex-}]45:fire}]  (unusedName) (\x1, 1.5);
			\end{pgfonlayer}
			
		\end{tikzpicture}
		\caption{Membrane potential curve}
		\label{fig:spiking_behaviour}
	\end{subfigure}\hfill
\begin{subfigure}{.6\linewidth}
	\centering
	\scalebox{.9}{%
	\begin{tikzpicture}[
		circuit logic US,
		]
		
		\node[circ] (leftsum) {$\sum$};
		\node[right=.5 of leftsum, box, fill=MembranePotentialSlopeColor] (ym) {$\mu_m$};
		\node[right=.5 of ym, circ] (rightsum) {$\sum$};
		\node[right=.5 of rightsum, memPot] (vm) {$V_m$};
		\node[below=.5 of rightsum, box, fill=white] (sawtooth) {
			\begin{tikzpicture}[scale=.7]
				\draw[ultra thick] (0,0) -- ++(.5,0) -- ++ (0,.3) -- ++ (.5,0);
			\end{tikzpicture}
		};
		\node[below left=.5 of sawtooth, box, fill=white] (tspike) {$t_{\text{spike}}$};
		\node[right=.5 of tspike, box, fill=white] (eqzero) {$= 0$};
		
		\node[right=.5 of sawtooth, and gate, draw,yshift=-4, fill=white] (and) {};

		\draw[thick] (and) -- coordinate[pos=1] (outputend) ++(1.5,0);
		\draw[spike] ($(and)!.8!(outputend)$) -- ++(0,.3);
		
		\node[box, rounded corners, left=of leftsum, fill=Greys-C, 
			] (w1) {$s_1$};
		\node[box, rounded corners, below=of w1, fill=Greys-C, 
			]  (w2) {$s_2$};
		
		\draw[thick] (w1.west) -- coordinate[pos=1] (w1end) ++(-1,0);
		\draw[thick] (w2.west) -- coordinate[pos=1] (w2end) ++(-1,0);
		
		\draw[spike] ($(w1)!.3!(w1end)$) -- ++(0,.3);
		\draw[spike] ($(w2)!.7!(w2end)$) -- ++(0,.3);

		\begin{scope}[
			every path/.style={draw, -latex, thick}
			]
			\draw (leftsum) -- (ym);
			\draw (ym) -- (rightsum);
			\draw (rightsum) -- (vm);
			\draw (ym.east) -- ($(ym.east)!.3!(rightsum.west)$) -- ++ (0,.75) node[outer sep=0pt] (helper1) {} --  (helper1 -| leftsum) -- (leftsum);
			\path (vm.east) -| ++(.25,-.5) node[outer sep=0pt] (helper3) {} -- (helper3 -| helper1) |- (sawtooth.west);
			\draw (sawtooth) -- coordinate[pos=.4] (sand) (sawtooth -| and.input 1);
			\draw (tspike) -- (eqzero);
			\draw (eqzero.east) -- (eqzero.east -| sand) |- (and.input 2);
			\draw (vm.east) -- ++ (.25,0) -- ++ (0,.75) node[outer sep=0pt] (helper2) {} --  (helper1 -| rightsum) -- (rightsum);
			
			\draw (w1) -- (leftsum);
			\draw (w2) -- (leftsum);
		\end{scope}

		\begin{pgfonlayer}{bg}
			\node[draw, circle, fit=(leftsum) (vm) (tspike) (helper1) (helper2), inner sep=0pt, fill=Greys-C, pin=45:TTFS neuron] (neuron) {};
		\end{pgfonlayer}
	\end{tikzpicture}
}
	\caption{TTFS neuron implementation }
	\label{fig:ttfs_neuron}
\end{subfigure}
\caption{Illustration of \subref{fig:spiking_behaviour}~the spiking behavior and \subref{fig:ttfs_neuron} implementation of a TTFS-encoded neuron as described in~\cite{rueckauer2}. In~\subref{fig:spiking_behaviour}, the membrane potential $V_m$ rises until reaching the firing threshold $V_t$. Then, a spike is emitted and the membrane potential is reset to zero. In~\subref{fig:ttfs_neuron}, the incoming spikes ($s_1, s_2$) are weighted ($w_1, w_2$) and added to the membrane potential slope $\mu_m$. The slope, in turn, causes the membrane potential to rise or fall. The value of the slope affects the rate of change of $V_m$.}
\label{fig:ttfs}
\end{figure}

On the other hand, in \ac{TTFS} encoding, not the firing rate of a neuron but the time it generates a spike for the first time is considered. 
The earlier this happens, the more important the spike is, \ie the higher the difference added to connected neurons' membrane potential is (see Figure~\ref{fig:ttfs}\subref{fig:spiking_behaviour} for an example).
The consequences of this are vastly increased processing speed for the evaluation of one neuron \cite{panzeri} and also for an entire \ac{SNN} as long as the sparsity of spikes is exploited.
Also, a neuron can only fire once, which leads to the fact that to reach acceptable accuracies using this method, \acp{SNN} need to be evaluated multiple times \cite{sommer}.

Figure~\ref{fig:ttfs}\subref{fig:ttfs_neuron} shows an implementation of a TTFS-neuron as described in~\cite{rueckauer2}. To ensure spike sparsity, neurons are only allowed to spike once. Hence, after emitting a spike, the neuron sets its internal $t_{\text{spike}}$ variable to $1$ prohibiting further spike emissions.
In this implementation, an additional value, the membrane potential slope $\mu_m$ is used to gain more fine-grained control over the rise and fall for the membrane potential $V_m$.

Using the slope $\mu_m$ incurs a large impact on the memory requirements of a neuron.
\citeauthor{Han2020}~\cite{Han2020} introduced a TTFS variant that does not use the slope $\mu_m$ and continuously emits spikes after reaching the membrane threshold $V_t$.
Following the notation introduced in~\cite{sommer}, we will call this variant m-TTFS.

\subsubsection{\ac{SNN} Training Methods}
Standard \acp{ANN} are most often trained using gradient-based backpropagation.
For this, all computational elements of the network must be differentiable.
For \acp{SNN}, the well-established backpropagation algorithm cannot be applied since spike events are inherently discontinuous and non-differentiable.
For this reason, other approaches have been proposed for \acp{SNN}.

Two major domains of techniques are: (a) training \acp{SNN} directly and (b) training a conventional \ac{ANN}, then converting it into an \ac{SNN} using the chosen spike encoding method.
Among (a), there are \ac{STDP} and spike-based backpropagation.

In \ac{STDP} \cite{stdp, stdp2}, synapses, \ie the connections between neurons \(i_1\) and \(i_2\) get assigned a higher weight whenever the time difference between spikes originating from \(i_1\) and \(i_2\) are small.
Lower weights are assigned if two neurons fire at different times.
Despite working directly on \acp{SNN}, achieving good accuracy using this method is difficult \cite{stdp2}.
The second technique based directly on \acp{SNN} is to use standard backpropagation but while using approximations or equivalent \ac{ANN} versions for the derivatives of \ac{SNN} operations, \ie membrane potential thresholding \cite{snnbackprop, snnbackprop2}.

The method used in more recent works is a conversion from a trained \ac{ANN} to \ac{SNN} (\eg \cite{sommer, syncnn}).
Here, a standard modeling framework like PyTorch can be used for training, which is a well-established technique.
Then, the trained weights are translated onto a \enquote{mirrored} \ac{SNN} architecture with corresponding layers \cite{rueckauer2,rueckauer}.
This process necessarily incurs an accuracy loss since conventional \ac{ANN} cannot incorporate the temporal dynamics of spiking systems, leading to degraded performance on event-based datasets \cite{snnbackprop}.
However, recent advances have improved the conversion error to below 0.4\%, even for the challenging ImageNet dataset \cite{rueckauer}. 

\subsection{Related Work}\label{sec:rel_work}
As has been done for traditional \acp{ANN}, extensive research work has been published regarding the hardware acceleration of \acp{SNN}.
Approaches can be compared considering several design objectives, including inference latency, required hardware resources, achieved classification accuracy, and energy requirements.
An extensive literature review reveals that most works use image classification datasets such as MNIST, SVHN, or CIFAR-10 for benchmarking and typically condense performance objectives into a single metric denoting energy efficiency in frames per second per Watt (FPS/W).
Further, these related works report only the average or maximal achievable frame rate and FPS/W, respectively.
In contrast, we show that the \ac{SNN}'s inference latency, and thus also corresponding energy, considerably depends on the input data. We, therefore, explicitly do not compute average values but show the full ranges instead.

Due to their closeness to biological neural models, \ac{SNN} research has produced approaches more concerned with flexibility, like the early SpiNNaker project, which is based on massively-parallel execution on ARM cores \cite{spinnaker}.
For a good overview of the field of \ac{SNN} accelerators, the interested reader is referred to the work by \citeauthor{snnsurvey}~\cite{snnsurvey}.

In the following, we review works most closely related to ours and categorize them into ASIC- and FPGA-based approaches.

\subsubsection{ASIC-based approaches}
Intel Loihi~\cite{loihi} is a chip design manufactured in a standard 14nm process implementing spiking neurons as an \ac{NoC}.
Spikes are represented as packets being sent in a unicast fashion between different neurons.
Advantages are an extremely high flexibility allowing inference and also training of a wide variety of \ac{SNN} architectures.
This comes at the cost of communication overhead, resulting in low energy efficiency.

IBM TrueNorth~\cite{truenorth} is a similar project based on an \ac{NoC} but restricted to tertiary weights, \ie values from the set \(\{-1, 0, 1\}\).
This leads to a very efficient design when it comes to power but reduced classification accuracy.

ASIE~\cite{asie} was proposed as an approach closely related to the work by \citeauthor{sommer}~\cite{sommer}, which we consider in the following, in that it encodes spikes as coordinates in a queue which is then processed until empty.
However, ASIE features a large array of \acp{PE}, which requires expensive routing and can lead to an under-utilization as it instantiates one \ac{PE} for each neuron in a layer, and layers differ in the numbers of neurons.

SNE~\cite{sne} is a highly-parallel ASIC design with compute engines arranged in an array for computing event-based convolutional layers.
It uses the leaky integrate-and-fire model and a spike encoding which includes neuron weights as well as the time of the event.
Spikes are distributed across the fixed-size array.
It has been evaluated using the N-MNIST dataset \cite{nmnist} for \acp{SNN} for which average energy efficiency numbers of more than 10.000~FPS/W are reported.

\subsubsection{FPGA-based approaches}
SIES~\cite{sies} is an accelerator designed explicitly for convolutional \acp{SNN} closely following  the architecture of traditional CNN accelerators.
The difference is that membrane potential changes can be calculated with only adders, requiring no \ac{MAC} operations.
With a \(64\times 64\) array of PEs, this, however, again does not exploit the spike sparsity in \acp{SNN} due to the spike encoding and fixed PE array.

\citeauthor{fang}~\cite{fang} propose an accelerator implemented using \ac{HLS} and standard MAC-based matrix multiplications but supporting temporal spike encoding.
This theoretically leads to a much lower classification latency but is quite expensive in terms of hardware resources and energy.

FireFly~\cite{firefly} is an accelerator design implemented in SpinalHDL~\cite{SpinalHDL}, featuring a \ac{PE} array for membrane potential updates.
A key advantage is the efficient usage of DSP resources for parallelized \emph{Multiplex}-Accumulate operations, yielding efficient resource usage on Xilinx UltraScale devices.
The training and deployment method via PyTorch~\cite{Paszke2019} and BrainCog~\cite{Zeng2022} leads to high flexibility but also reduced efficiency for specific workloads like MNIST evaluation.

SyncNN~\cite{syncnn} is an HLS-based implementation of a queue-processing accelerator involving mixed-precision quantization and several other hardware optimizations.
It achieves a very high energy efficiency on various datasets and can be synthesized for different network models.
In SyncNN, spikes are represented not as binary values but as numbers representing how often a neuron has spiked.
These values are then multiplied together with kernel weights to produce membrane potential slopes.
As such, it can be regarded as a hybrid approach that sequentially processes layers using multiplications but with sparse and very low-precision activations.

Cerebron~\cite{cerebron} is an FPGA-based accelerator for SNNs that uses a systolic array.
Its specialty is its support for depthwise separable convolutions where a single filter output can be broadcast to multiple compute units. 
This reduces the memory requirements and improves the energy efficiency as long as suitable network architectures and training methods are used.
However, a significant scheduling overhead is involved in gaining these advantages in addition to suffering from an increased hardware complexity.

An approach using \ac{STDP} as its training method, instead of conversion from \acp{CNN} to \acp{SNN}, is Spiker~\cite{spiker}.
It implements an MNIST classifier using a single layer only. This results in a relatively low accuracy of 77\%.
Moreover, the design cannot be easily adapted to deeper networks or other datasets.

\citeauthor{gyro}~\cite{gyro} present an FPGA-based accelerator for \acp{SNN} whose Gyro architecture is restricted to fully connected layers arranged in a pipeline with weight memories in between.
The \ac{SNN} architecture has exclusively been evaluated for the specific task of pixel-wise farmland classification into different types of crops using fused optical-radar data -- no results for other benchmarks (datasets), especially the commonly used benchmarks MNIST, SVHN or CIFAR-10, are reported.
Further, key performance indicators are provided as a function of the number of synaptic operations; therefore, the approach can hardly be numerically compared  with other accelerators.

An executive summary of related works discussed in this section with respect to target platforms and spike encoding schemes is given in Table~\ref{tbl:related}.

\pgfplotstableread[col sep=&, header=true]{./tables/relWork.tbl}{\relWorkTable}

\pgfplotstableset{
	columns/Work/.style={string type, column type={l}},
	columns/Platform/.style={string type},
	columns/Spike Encoding/.style={string type},}

\begin{table}
	\caption{Overview of existing SNN implementations with respect to the used technology (ASIC/FPGA) and used neuron model.}
	\label{tbl:related}
	\pgfplotstabletypeset[
		columns/Work/.style={string type,column type={l}},
		columns/Platform/.style={string type},
		columns/Spike Encoding/.style={string type},
		columns={Work, Platform, Spike Encoding},
		every row no 3/.style={after row=\midrule},
        skip rows between index={11}{21}
	]{\relWorkTable}

\end{table}

ASIC designs (such as Loihi~\cite{loihi}) tend to model the biologically inspired features of \acp{SNN} better.
	The FPGA-based implementations as considered in this paper are accelerators exploiting the sparsity in \ac{SNN} forward computations but are less biologically inspired.
	Because of the reprogrammability of FPGAs, designs can be tailored for a specific network.
	This, in connection with the manufacturing process, makes ASICs less comparable to the FPGA-based designs, which we focus on in the following.


\section{Neural Network Hardware Architectures}\label{sec:archs}
In the following, we present the fundamental \ac{SNN} hardware architecture used in our work (Section~\ref{sec:SNNarch}) as well as the basic architectural concepts of CNN accelerators (Section~\ref{sec:CNNarch}) employed throughout our quantitative comparison.
\subsection{Spiking Neural Network Architecture\label{sec:SNNarch}}
\begin{figure}
\begin{tikzpicture}[
		large/.style={minimum height=4cm}
	]
	\node[aeq, large] (aeq) {AEQ};
	\node[genBox, right=of aeq, minimum width=60pt] (convU) {Convolution\\ Unit};
	\node[rom, below=.57 of convU, minimum width=60pt] (kernelROM) {Kernel \\ ROM};
	\node[memPot, right=of convU, large] (memPot) {Mem\\Pot};
	\node[genBox, right=of memPot, minimum width=60pt] (threshU) {Thresholding\\ Unit};
	\node[rom, below=.57 of threshU, minimum width=60pt] (biasROM) {Bias\\ ROM};
	\node[classU, above=.57 of convU, minimum width=60pt] (classU) {Classification\\ Unit};
	
	\begin{scope}[
		every path/.style={draw, -latex, thick}
	]
		\draw (aeq) -- coordinate[pos=.4] (fanout) (convU);
		\draw (fanout) |- (classU);
		\draw  ($(convU.east)+(0,.2)$) -- ($(memPot.west)+(0,.2)$);
		\draw[latex-]  ($(convU.east)+(0,-.2)$) -- ($(memPot.west)+(0,-.2)$);
		\draw (biasROM) -- (threshU);
		\draw (kernelROM) -- (convU);
		\draw ($(memPot.east)+(0,.2)$) -- ($(threshU.west)+(0,.2)$);
		\draw[latex-] ($(memPot.east)+(0,-.2)$) -- ($(threshU.west)+(0,-.2)$);
		\draw (threshU.east) -- ++(0.25,0) -- ++(0,2.25) -- ++ (-10,0) |- (aeq.west);
	\end{scope}
\end{tikzpicture} 
\caption{Overview of the SNN accelerator architecture, as proposed by~\citeauthor{sommer}~\cite{sommer}. The incoming spikes are stored in the queue \ac{AEQ} (blue), and the membrane potentials are stored in the queue Mem Pot (green). After all spikes in the queue have been processed, newly emitted spikes are fed into the \ac{AEQ}, which is empty again.}
\label{fig:sommerarch}
\end{figure}
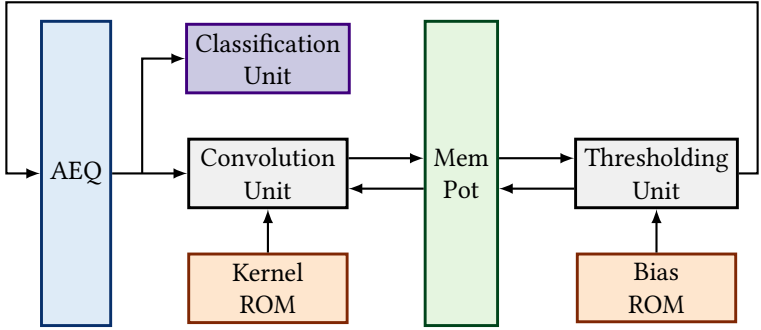

For our comparative analysis, we investigate a recently published, state-of-the-art work: the unnamed approach of~\citeauthor{sommer}\footnote{We thank the authors for providing us access to the \ac{SNN} accelerator's VHDL code.}~\cite{sommer}.
This architecture is chosen as it exploits the sparsity in \acp{SNN} as well as multiplier-less implementations of convolutional layers as concepts to achieve high expected energy efficiency.
It also features a high degree of configurability, allowing us to match resource usage and frequency on a given platform and measure the resulting changes (see Figure~\ref{fig:sommerarch} for an overview of the architecture).

As the accelerator targets convolution operations (e.g., in image classification tasks), its design centers around two-dimensional matrices: the spatial arrangement of the incoming spikes (called a \emph{feature map}) and the kernel matrix used in the convolution. 
Consequently, spikes are understood as events associated with a location within the feature map and are consequently named \acfp{AE}.
These events are then stored in \acfp{AEQ} that allow processing them in order.

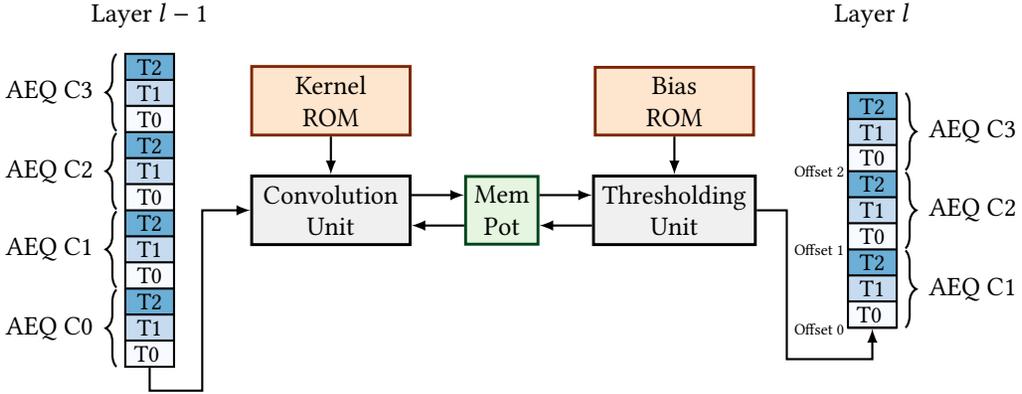
\begin{figure}
	\begin{tikzpicture}[
			aeql/.style={
				rectangle split,
				rectangle split allocate boxes=12,
				rectangle split parts=12,
				rectangle split ignore empty parts=true,
				rectangle split part fill={Blues-G, Blues-D,Blues-A,Blues-G, Blues-D,Blues-A,Blues-G, Blues-D,Blues-A,Blues-G, Blues-D,Blues-A,},
				draw=black,
				align=center,
				thick,
				inner ysep=1.5pt,
				font=\small
			},
			curly/.style n args={1}{
				decorate,decoration={brace,amplitude=5pt, #1, raise=2pt}, thick, shorten <=1pt, shorten >=1pt
			}
		]
		
		\node[memPot] (memPot) {Mem\\ Pot};
		\begin{scope}[
				every node/.style={
					align=center,
					minimum width=60pt,
				}
			]

			\node[genBox, left=.7 of memPot] (convU) {Convolution\\ Unit};
			\node[genBox, right=.7 of memPot] (threshU) {Thresholding\\ Unit};
			\node[rom, above=.5 of convU] (kernelROM) {Kernel \\ ROM};
			\node[rom, above=.5 of threshU] (biasROM) {Bias\\ ROM};
		\end{scope}

		\node[aeql, left=of convU] (leftColumn) {
			\nodepart{one}
			T2
			\nodepart{two}
			T1
			\nodepart{three}
			T0
			\nodepart{four}
			T2
			\nodepart{five}
			T1
			\nodepart{six}
			T0
			\nodepart{seven}
			T2
			\nodepart{eight}
			T1
			\nodepart{nine}
			T0
			\nodepart{ten}
			T2
			\nodepart{eleven}
			T1
			\nodepart{twelve}
			T0
		};
		\node[above=.2 of leftColumn, align=center] (lastLabel) {Layer $l-1$};
				\begin{scope}[
			every path/.style={curly=mirror},
			every node/.style={midway, left=8pt}
			]
			\draw (leftColumn.north west) -- (leftColumn.three split west) node {AEQ C3};			
			\draw (leftColumn.three split west) -- (leftColumn.six split west) node {AEQ C2};
			\draw (leftColumn.six split west) -- (leftColumn.nine split west) node {AEQ C1};
			\draw (leftColumn.nine split west) -- (leftColumn.south west) node {AEQ C0};
		\end{scope}
		
		\node[aeql, right=1.2 of threshU] (rightColumn) {
			\nodepart{one}
			T2
			\nodepart{two}
			T1
			\nodepart{three}
			T0
			\nodepart{four}
			T2
			\nodepart{five}
			T1
			\nodepart{six}
			T0
			\nodepart{seven}
			T2
			\nodepart{eight}
			T1
			\nodepart{nine}
			T0
		};
		\begin{scope}[
			every path/.style={curly},
			every node/.style={midway, right=8pt}
			]
			\draw (rightColumn.north east) -- (rightColumn.three split east) node {AEQ C3};			
			\draw (rightColumn.three split east) -- (rightColumn.six split east) node {AEQ C2};
			\draw (rightColumn.six split east) -- (rightColumn.south east) node {AEQ C1};
		\end{scope}
	
		\node[align=center] (currLabel) at (lastLabel -| rightColumn.center) {Layer $l$};

		\coordinate (helper) at ($(rightColumn.south)+(0,-.4)$);

		\begin{scope}[
			every path/.style={draw, -latex, thick}
			]
			\draw  ($(convU.east)+(0,.2)$) -- ($(memPot.west)+(0,.2)$);
			\draw[latex-]  ($(convU.east)+(0,-.2)$) -- ($(memPot.west)+(0,-.2)$);
			\draw (biasROM) -- (threshU);
			\draw (kernelROM) -- (convU);
			\draw ($(memPot.east)+(0,.2)$) -- ($(threshU.west)+(0,.2)$);
			\draw[latex-] ($(memPot.east)+(0,-.2)$) -- ($(threshU.west)+(0,-.2)$);
			\draw (leftColumn.south) -- ++(0,-.3) -- ++ (.7,0) |- (convU.west);
			\draw (threshU.east) -- +(.4,0) coordinate (h) -- (h |- helper) -- (helper) -- (rightColumn);
		\end{scope}
	
		\node[anchor=east, font=\tiny, outer sep=1pt, inner sep=0pt] at (rightColumn.three split west) {Offset 2};
		\node[anchor=east, font=\tiny, outer sep=1pt, inner sep=0pt] at (rightColumn.six split west) {Offset 1};
		\node[anchor=east, font=\tiny, outer sep=1pt, inner sep=0pt] at (rightColumn.south west) {Offset 0};
	\end{tikzpicture}
	\caption{Visualization of the use and segmentation of \acp{AEQ} as spike storage. \(T_i\) refers to the algorithmic time step, while \(C_i\) are the input and output channels of the convolutional layer. }
	\label{fig:tick-wise-processing}
\end{figure}

FPGA memory resources (BRAM or LUTRAM) can be used to implement these AEQs.
Using an addressing scheme that divides these memories into segments depending on the algorithmic time step \(t\), input and output channel, and layer, they allow one kernel operation in a convolutional layer to be processed at a time.
That is, loading of the membrane potentials of a neuron together with its neighborhood is achieved within one clock cycle.
This is visualized in Figure~\ref{fig:tick-wise-processing}.
\acp{AEQ} are basically a two-dimensional array of spike arrays, with the first dimension being channels of the convolutional layer, and the second corresponding to algorithmic time steps.

By pipelining the computations, a throughput of one spike per cycle per core can be achieved as long as the queues are filled.
By replicating these cores, and distributing spike events across them, the whole spike processing can be parallelized.
Here, parallelization factors between 1 and 16 have been tested.

\newcommand{\memNode}[3]{\indexedMatrixNodeHelper{memPotMatrixNode}{#1}{#2}{#3}}
\newcommand{\aeqNode}[3]{\indexedMatrixNodeHelper{aeqMatrixNode}{#1}{#2}{#3}}
\newcommand{\indexedMatrixNodeHelper}[4]{%
	\node[#1] (#1-\thememPotNodeCounter) {#2};%
	\node[red, above right=-.3 and -.2 of #1-\thememPotNodeCounter, inner sep=0pt, outer sep=0pt] {\bfseries\small #3};
	\node[below right=-.32 and -.65 of #1-\thememPotNodeCounter, inner sep=0pt, outer sep=0pt] {\bfseries\small #4};
	\stepcounter{memPotNodeCounter}
}
\newcommand{\mementryhelper}[7][]{
	\matrix[#7, #1, outer sep=0pt, inner sep=0pt] (mem-#2-M) {
		#3\\
		#4\\
		#5\\
		#6\\
	};
	\node[above = 1pt of mem-#2-M] (mem-#2-M-top) {\bfseries \textcolor{red} #2};
}
\newcommand{\imementry}[6][]{\mementryhelper[#1]{#2}{#3}{#4}{#5}{#6}{imem}}
\newcommand{\aeqmementry}[6][]{\mementryhelper[#1]{#2}{#3}{#4}{#5}{#6}{aeqMem}}

\newcounter{memPotNodeCounter}

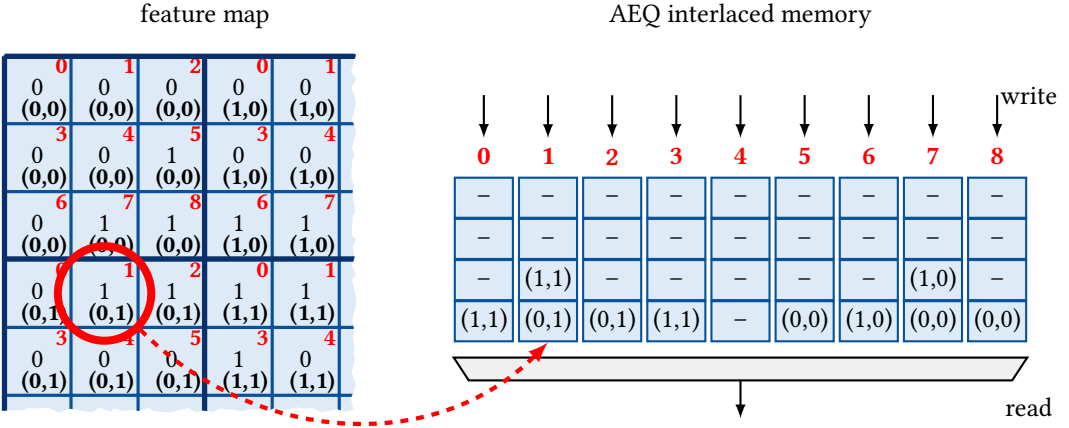
\begin{figure}
\begin{tikzpicture}
	\setcounter{memPotNodeCounter}{0}
	\begin{scope}
		\clip node [
		matrix, 
		anchor=center, 
		fancyMatrix,
		append after command={%
			coordinate (start) at ($(aeqMatrixNode-0.north west)+(-.2,.2)$)
			coordinate (h1) at ($(aeqMatrixNode-4.north east)+(.2,.2)$)%
			coordinate (h2) at ($(aeqMatrixNode-28.south)+(0,-.2)$)%
			coordinate (h3) at ($(aeqMatrixNode-24.south west)+(-.2,.-2)$)
			(start) -- (h1) decorate[decoration={random steps,segment length=3pt,amplitude=1pt}] {-- (h2 -| h1) -- (h3 |- h2)} -- cycle%
		}%
		] (fmap) {
			\aeqNode{0}{0}{(0,0)} & \aeqNode{0}{1}{(0,0)} & \aeqNode{0}{2}{(0,0)} & \aeqNode{0}{0}{(1,0)} & \aeqNode{0}{1}{(1,0)} & \aeqNode{0}{4}{(1,1)}\\
			\aeqNode{0}{3}{(0,0)} & \aeqNode{0}{4}{(0,0)} & \aeqNode{1}{5}{(0,0)} & \aeqNode{0}{3}{(1,0)} & \aeqNode{0}{4}{(1,0)} & \aeqNode{0}{4}{(1,1)}\\
			\aeqNode{0}{6}{(0,0)} & \aeqNode{1}{7}{(0,0)} & \aeqNode{1}{8}{(0,0)} & \aeqNode{1}{6}{(1,0)} & \aeqNode{1}{7}{(1,0)} & \aeqNode{0}{4}{(1,1)}\\
			\aeqNode{0}{0}{(0,1)} & \aeqNode{1}{1}{(0,1)} & \aeqNode{1}{2}{(0,1)} & \aeqNode{1}{0}{(1,1)} & \aeqNode{1}{1}{(1,1)} & \aeqNode{0}{4}{(1,1)}\\
			\aeqNode{0}{3}{(0,1)} & \aeqNode{0}{4}{(0,1)} & \aeqNode{0}{5}{(0,1)} & \aeqNode{1}{3}{(1,1)} & \aeqNode{0}{4}{(1,1)} & \aeqNode{0}{4}{(1,1)}\\
			\aeqNode{0}{}{(0,1)} & \aeqNode{0}{}{(0,1)} & \aeqNode{0}{}{(0,1)} & \aeqNode{1}{}{(1,1)} & \aeqNode{0}{}{(1,1)} & \aeqNode{0}{}{(1,1)}\\
		};
		\draw[line width=2pt, draw=AEQDark] (aeqMatrixNode-2.north east) -- (aeqMatrixNode-32.south east);
		\draw[line width=2pt, draw=AEQDark] (aeqMatrixNode-12.south west) -- (aeqMatrixNode-17.south east);
		\draw[line width=2pt, draw=AEQDark] (aeqMatrixNode-0.north west) -- (aeqMatrixNode-5.north east);
		\draw[line width=2pt, draw=AEQDark] (aeqMatrixNode-0.north west) -- (aeqMatrixNode-30.south west);
	\end{scope}
	\node[above=3pt of fmap] (fmapLabel) {feature map};
	
	\aeqmementry[right=.5cm of fmap]{0}{--}{--}{--}{(1,1)}
	\aeqmementry[right=1.5pt of mem-0-M]{1}{--}{--}{(1,1)}{(0,1)}
	\aeqmementry[right=1.5pt of mem-1-M]{2}{--}{--}{--}{(0,1)}
	\aeqmementry[right=1.5pt of mem-2-M]{3}{--}{--}{--}{(1,1)}
	\aeqmementry[right=1.5pt of mem-3-M]{4}{--}{--}{--}{--}
	\aeqmementry[right=1.5pt of mem-4-M]{5}{--}{--}{--}{(0,0)}
	\aeqmementry[right=1.5pt of mem-5-M]{6}{--}{--}{--}{(1,0)}
	\aeqmementry[right=1.5pt of mem-6-M]{7}{--}{--}{(1,0)}{(0,0)}
	\aeqmementry[right=1.5pt of mem-7-M]{8}{--}{--}{--}{(0,0)}
	
	\foreach \n in {0,...,8} \draw[latex-, thick] (mem-\n-M-top) -- coordinate[pos=1] (mem-in-\n)  ++(0,.8);
	\node[xshift=1.2em] (writeLabel) at (mem-in-8) {write};

	\node (interlacedLabel )at (fmapLabel -| mem-4-M-1-1) {AEQ interlaced memory};
	
	\draw[thick, draw=GenericDark, fill=GenericColor] ($(mem-0-M.south west)+(0,-.2)$) -- ($(mem-8-M.south east)+(0,-.2)$) -- ++ (-.2,-.3) -- coordinate[pos=.5] (readOut) ($(mem-0-M.south west)+(.2,-.5)$) -- cycle;
	
	\draw[thick,-latex] (readOut) -- coordinate[pos=.7] (helper) ++(0,-.5);
	
	\node (readLabel) at (helper -| writeLabel) {read};

	\node[fit=(aeqMatrixNode-19), draw=red, line width=3pt, circle, inner sep=0pt] (marker) {};
	\draw[dashed, red, ultra thick, -latex] (marker) to[in=-135,out=-45] (mem-1-M.south);
\end{tikzpicture}
\vspace{-2em}
\caption{Memory interlacing for \acfp{AEQ}: The highlighted input spike is at position \textbf{\textcolor{red}{\textrm{1}}} in the kernel coordinate system (indicated by red numbers in the feature map). Hence, the value is put into queue number 1. Its value in the address coordinate system is (0,1), as indicated by the tuples in the feature map.
}
\label{fig:aeqinterlacing}
\end{figure}

Membrane potentials must also be stored for each neuron but only twice for one layer at a time (see block MemPot in Figures~\ref{fig:sommerarch} and~\ref{fig:tick-wise-processing}).
This number is sufficient as \acp{SNN} are processed one layer at a time.
The duplication is due to the thresholding (see Eq.~(\ref{eq:ifx})) being performed as a separate step after computing the new memory potentials.
A double buffering strategy is therefore used to pipeline these operations: Thresholding of one feature map is done while for the next map, new memory potentials are already computed.
The Tresholding Unit also computes and encodes new address events (\ie spikes) into the queues to be processed once the next layer is scheduled.

Since only one word can be read simultaneously from physical memory, a memory interlacing scheme is used to parallelize the access to both the feature map (storing the spikes still to be processed) and the membrane potential.
To perform a convolution at a given point of a feature map, the neighborhood of neurons, as defined by the kernel size, needs to be accessed.
This means that for a \(3\times 3\) kernel, $9$ neurons need to be checked for an incoming spike. To allow for a parallel access to these neurons, memory resources are replicated nine times to increase throughput. The idea is to divide the feature map into windows of kernel size, resulting in a coarser grid of coordinates, or \emph{addresses}, $(x,y)$ than before. Within each window, the individual neurons are enumerated from $0$ to the size of the window minus one (we will call this the kernel coordinate system). The spikes of the feature map are then stored in an \acf{AEQ} as follows. The \ac{AEQ} consists of as many BRAM-based queues as the kernel size ($3\times 3=9$ in our example). The kernel and address coordinate system uniquely identify all spikes. The address of each spike is stored in the queue corresponding to its kernel coordinate system value.
See Figure~\ref{fig:aeqinterlacing} for an illustration of this principle.

\begin{figure}
\begin{tikzpicture}

	\setcounter{memPotNodeCounter}{0}
	\begin{scope}
	\clip node [
		matrix, 
		anchor=center, 
		fancyMatrix,
		append after command={%
			coordinate (start) at ($(memPotMatrixNode-0.north west)+(-.2,.2)$)
			coordinate (h1) at ($(memPotMatrixNode-4.north east)+(.2,.2)$)%
			coordinate (h2) at ($(memPotMatrixNode-28.south)+(0,-.2)$)%
			coordinate (h3) at ($(memPotMatrixNode-24.south west)+(-.2,.-2)$)
		 	(start) -- (h1) decorate[decoration={random steps,segment length=3pt,amplitude=1pt}] {-- (h2 -| h1) -- (h3 |- h2)} -- cycle%
		}%
	] (memPotM) {
		\memNode{2}{0}{(0,0)} & \memNode{2}{1}{(0,0)} & \memNode{3}{2}{(0,0)} & \memNode{0}{0}{(1,0)} & \memNode{0}{1}{(1,0)} & \memNode{0}{1}{(1,0)}\\
		\memNode{1}{3}{(0,0)} & \memNode{3}{4}{(0,0)} & \memNode{7}{5}{(0,0)} & \memNode{3}{3}{(1,0)} & \memNode{3}{4}{(1,0)} & \memNode{0}{1}{(1,0)}\\
		\memNode{0}{6}{(0,0)} & \memNode{1}{7}{(0,0)} & \memNode{4}{8}{(0,0)} & \memNode{5}{6}{(1,0)} & \memNode{5}{7}{(1,0)} & \memNode{0}{1}{(1,0)}\\
		\memNode{0}{0}{(0,1)} & \memNode{0}{1}{(0,1)} & \memNode{0}{2}{(0,1)} & \memNode{3}{0}{(1,1)} & \memNode{3}{1}{(1,1)} & \memNode{0}{1}{(1,0)}\\
		\memNode{0}{3}{(0,1)} & \memNode{0}{4}{(0,1)} & \memNode{0}{5}{(0,1)} & \memNode{3}{3}{(1,1)} & \memNode{3}{4}{(1,1)} & \memNode{0}{1}{(1,0)}\\
		\memNode{0}{}{(0,1)} & \memNode{0}{}{(0,1)} & \memNode{0}{}{(0,1)} & \memNode{3}{}{(1,1)} & \memNode{3}{}{(1,1)} & \memNode{0}{}{(1,0)}\\
		};
		\draw[line width=2pt, draw=MembranePotentialDark] (memPotMatrixNode-2.north east) -- (memPotMatrixNode-32.south east);
		\draw[line width=2pt, draw=MembranePotentialDark] (memPotMatrixNode-12.south west) -- (memPotMatrixNode-17.south east);
		\draw[line width=2pt, draw=MembranePotentialDark] (memPotMatrixNode-0.north west) -- (memPotMatrixNode-5.north east);
		\draw[line width=2pt, draw=MembranePotentialDark] (memPotMatrixNode-0.north west) -- (memPotMatrixNode-30.south west);
	\end{scope}
	\draw[red, ultra thick] (memPotMatrixNode-8.north west) -- (memPotMatrixNode-10.north east) -- (memPotMatrixNode-22.south east)-- (memPotMatrixNode-20.south west) -- cycle ;

	\node[above=3pt of memPotM] (memPotLabel) {MemPot};

	\imementry[right=1.0cm of memPotM]{0}{2}{0}{0}{\textcolor{red}3}
	\imementry[right=1.5pt of mem-0-M]{1}{2}{0}{0}{\textcolor{red}3}
	\imementry[right=1.5pt of mem-1-M]{2}{3}{\textcolor{red}0}{0}{0}
	\imementry[right=1.5pt of mem-2-M]{3}{1}{0}{\textcolor{red}3}{3}
	\imementry[right=1.5pt of mem-3-M]{4}{3}{0}{\textcolor{red}3}{3}
	\imementry[right=1.5pt of mem-4-M]{5}{\textcolor{red}7}{0}{0}{0}
	\imementry[right=1.5pt of mem-5-M]{6}{0}{0}{\textcolor{red}5}{0}
	\imementry[right=1.5pt of mem-6-M]{7}{1}{0}{\textcolor{red}5}{0}
	\imementry[right=1.5pt of mem-7-M]{8}{\textcolor{red}4}{0}{0}{0}
	
	\node[left=1.5pt of mem-0-M-1-1, outer sep=0pt, inner sep=0pt] {(0,0)};
	\node[left=1.5pt of mem-0-M-2-1, outer sep=0pt, inner sep=0pt] {(0,1)};
	\node[left=1.5pt of mem-0-M-3-1, outer sep=0pt, inner sep=0pt] {(1,0)};
	\node[left=1.5pt of mem-0-M-4-1, outer sep=0pt, inner sep=0pt] {(1,1)};
	
	\node (interlacedLabel )at (memPotLabel -| mem-4-M-1-1) {MemPot interlaced memory};

	\begin{scope}[every node/.style={imemNode,draw=red, fill=none, ultra thick}]
		\node at (mem-0-M-4-1) {\phantom{0}};
		\node at (mem-1-M-4-1) {\phantom{0}};
		\node at (mem-2-M-2-1) {\phantom{0}};
		\node at (mem-3-M-3-1) {\phantom{0}};
		\node at (mem-4-M-3-1) {\phantom{0}};
		\node at (mem-5-M-1-1) {\phantom{0}};
		\node at (mem-6-M-3-1) {\phantom{0}};
		\node at (mem-7-M-3-1) {\phantom{0}};
		\node at (mem-8-M-1-1) {\phantom{0}};
	\end{scope}
	
	\node[draw=red, circle, line width=3pt, fit=(memPotMatrixNode-22), inner sep=0pt] (marker) {};
	\draw[red, dashed, ultra thick, -latex] (marker) to[out=-45,in=225] (mem-1-M-4-1.south);
\end{tikzpicture}
\vspace{-1.5em}
	\caption{Memory interlacing for membrane potentials. Any placement of the kernel is guaranteed to select exactly one neuron per memory. The neurons selected by the kernel indicated by the red square on the left are highlighted in red on the right. One can see that exactly one value per memory needs to be retrieved.}
	\label{fig:sommerinterlacing}
\end{figure}
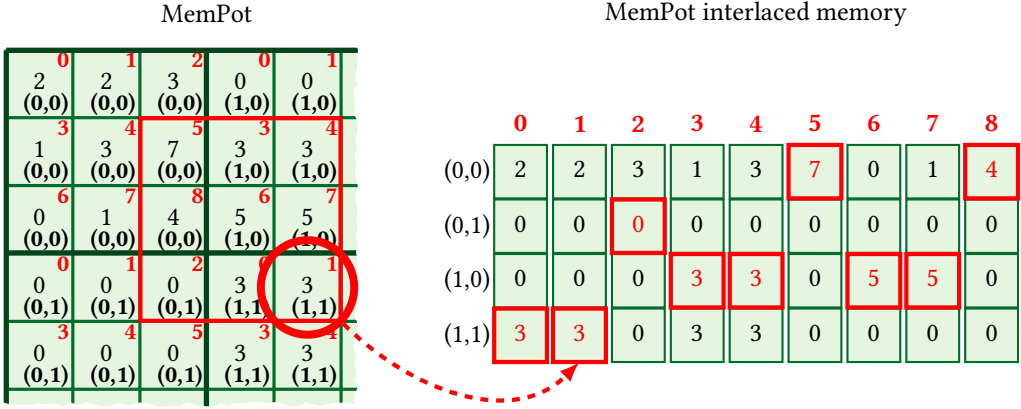

For the membrane potentials, a similar interlacing scheme is applied. Instead of storing the addresses within the individual queues, they are used to define the memory depth $D$.  In a single convolution step, all membrane potentials of neurons in the kernel neighborhood need to be retrieved. The kernel and address coordinates combined allow to uniquely identify the memory potential of any neuron. Furthermore, the addressing/interlacing scheme guarantees that no concurrent read access to the memories are carried out (see Figure~\ref{fig:sommerinterlacing} for a visualization).

The approach follows the method of converting a trained traditional \ac{CNN} to an \ac{SNN}.
In~\cite{sommer}, \texttt{snntoolbox}~\cite{rueckauer} is used for this purpose.
As a result, accuracy drops of less than 0.4\% can be achieved when comparing the converted net to the original \ac{CNN} for the MNIST benchmark.


\subsection{Convolutional Neural Network Architectures\label{sec:CNNarch}}
Hardware implementations for \acp{CNN} have been proposed in a large number for which many surveys and overviews exist in the literature, \eg \cite{safari, annreview}.
Currently, the state-of-the-art does not consist of single configurable accelerators (probably specifically tuned for certain use cases) but entire compiler toolchains such as FINN~\cite{Blott2018}. It should be noted, though, that many are still a work in progress.
A compiler-based approach can transform an input \ac{NN} into a hardware design ready to be deployed on an FPGA.
In the survey by \citeauthor{safari}~\cite{safari}, a differentiation between \textit{overlay-based} and \textit{dedicated} accelerators has been introduced.
The former uses a fixed kernel and creates an instruction stream to pipe data through this/these kernels.
By contrast, a dedicated accelerator implements the entire network directly in hardware, including weights, quantization, or pruning settings.
In the following, we focus on dedicated hardware accelerators for \acp{CNN} since these have not yet been considered for comparison with \ac{SNN} accelerators.


\colorlet{FINNblue}{Blues-C}
\colorlet{FINNgreen}{Greens-D}

\newcommand{\boxes}[1][.3]{%
	\begin{tikzpicture}
		\begin{scope}[scale=#1, anchor=center]
		\draw[fill=FINNblue, xshift=-8, yshift=8] (0,0) rectangle (1,1);
		\draw[fill=FINNblue, xshift=-5, yshift=5] (0,0) rectangle (1,1);
		\draw[fill=FINNblue, xshift=-2, yshift=2] (0,0) rectangle (1,1);
		\end{scope}
	\end{tikzpicture}
}
\begin{figure}
\begin{tikzpicture}[
	box/.style={align=center, draw, fill=FINNgreen, text depth=1cm},
	inlet/.style={align=left, draw, fill=FINNblue, anchor=south}
	]

	\node[box] (layer1) {layer 1\\compute array};
	\node[inlet] (weights1) at (layer1.south) {\boxes weights};

	\node[box, right=of layer1] (layer2) {layer 2\\compute array};
	\node[inlet] (weights2) at (layer2.south) {\boxes weights};

	\node[box, right=of layer2] (layer3) {layer $L$\\compute array};
	\node[inlet] (weights3) at (layer3.south) {\boxes weights};

	\node[left=2cm of layer1, anchor=center, rotate=90] (onchip) {\bfseries on-chip};

	\node[below left=2cm and 1cm of layer1,  label={[label distance=.01cm]35:images}] (images) {\boxes[.5]};
	\node[below right=2cm and 1cm of layer3, label distance=.1, label={[label distance=.01cm]145:classifications}] (classifications) {\boxes[.5]};
	\node[anchor=center, rotate=90] (offchip) at (onchip |- images) {\bfseries off-chip};

	\node[above=of layer2] (labelAbove) {heterogeneously sliced; tailored to compute requirements};
	\node[below=.4 of layer2] (labelBelow) {carry intermediate activations via on-chip channels (FIFOs)};

	\begin{scope}[every path/.style={dashed, -latex}]
		\draw (layer1) to[out=90,in=-90] (labelAbove);
		\draw (layer2) -- (labelAbove);
		\draw (layer3) to[out=90, in=-90] (labelAbove);
	\end{scope}

	\begin{scope}[every path/.style={draw, -latex, thick}]
		\draw (images) |- (layer1.west);
		\draw (layer1) -- coordinate (i1) (layer2);
		\draw (layer2) -- coordinate (i2) (layer3);
		\draw (layer3.east) -| (classifications);
	\end{scope}

	\node[below=.1cm of i1, outer sep=0pt] (iboxes1) {\boxes};
	\node[below=.1cm of i2, outer sep=0pt] (iboxes2) {\boxes};

	\node[above=.1cm of i2, outer sep=0pt] (dots) {\bfseries ...};

	\draw[dashed, thick, shorten >=2pt] (iboxes1) -- (labelBelow -| iboxes1);
	\draw[dashed, thick, shorten >=3pt] (iboxes2) -- (labelBelow -| iboxes2);

	\node[draw, cloud, cloud ignores aspect, cloud puffs=14, cloud puff arc=90,  text width=3cm, yshift=-.5cm, fill=FINNblue] at (images -| layer2) {External memory or peripheral devices};

	\coordinate (h) at ($(onchip)!.65!(offchip)$);
	\draw[dotted, ultra thick] (h) -- (h -| classifications.east);
\end{tikzpicture}
\caption{FINN-generated CNN architecture. Reproduced from~\cite[Fig. 10]{Blott2018}.}
\label{fig:finnlayers}
\end{figure}
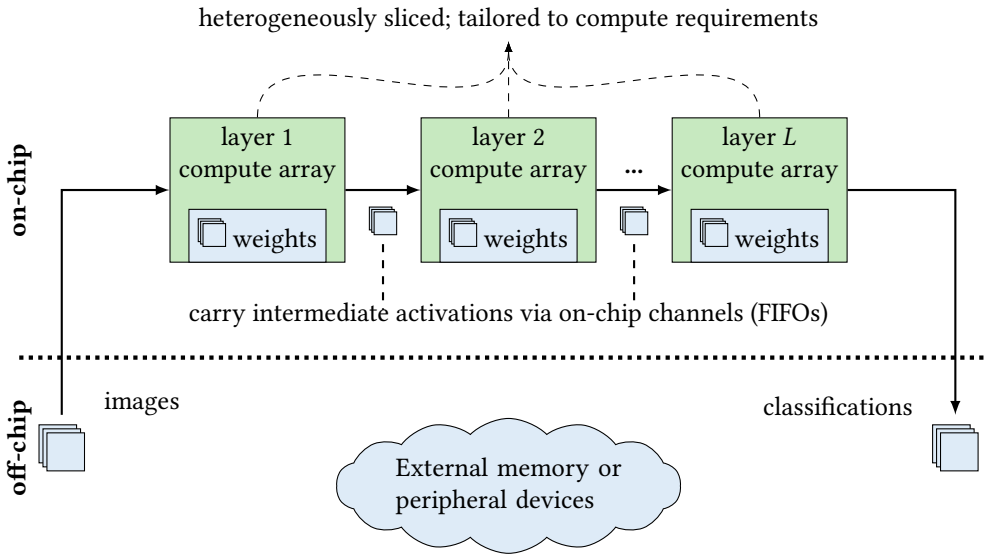

We employ the FINN framework \cite{Blott2018} for our comparative analysis to generate efficient, dedicated \ac{CNN} accelerators.
FINN creates so-called streaming dataflow architectures where each layer is instantiated as an array of \acp{PE} with FIFO buffers in between (see Figure~\ref{fig:finnlayers}).
In FINN accelerators, all network layers execute in a parallel way.
The complete computation is pipelined with layers being implemented as IP cores connected using self-synchronizing protocols and FIFOs in between for storing intermediate results.

A central concept in FINN is the use of mixed-precision quantization techniques for reducing memory and resource usage.
This is accomplished by the accompanying Brevitas tool, a PyTorch library, which can export \acp{NN} in the FINN-readable ONNX format.
Basically, the FINN operators are combined into hardware-suitable custom operators.
These operators are then mapped to instantiations of HLS modules with corresponding configurations, which are connected together in the Xilinx Vivado toolchain.
Finally, synthesis and global place and route can be performed on this input to produce an FPGA configuration.

Figure~\ref{fig:finnlayers} illustrates how the CNN layers are connected in a hardware architecture generated by FINN and how weights and activations are stored.
As \acp{PE}, FINN uses \ac{MAC} units in combination with adder trees to implement matrix-vector multiplications.
This is sufficient to process fully connected layers.
Convolutional layers are mapped to a so-called \emph{sliding window unit}, which buffers and re-orders the input feature map so the \ac{MAC} units can seamlessly be reused.
Only several rows of the feature map need to be buffered at a time, depending on kernel size.
Likewise, weights need to be kept in memory in full for all layers and channels, as depicted.

The configuration of the \ac{MAC} units is the core deciding factor regarding resource usage and latency of the resulting design.
Each \ac{PE} computes \(Q_l\) multiplications in parallel (SIMD value), and \(P_l\) \acp{PE} are instantiated for layer \(l\).
This tends to reduce latency and increase resource usage linearly depending on \(Q_l \cdot P_l\).
However, as the whole network is executed as a pipeline, the layer whose configuration least matches its compute intensity (\ie low \(P_l\) and \(Q_l\) while needing many multiplications) limits the throughput.

\section{Experimental Results}
\label{sec:experiments}
In this section, we compare \ac{SNN} and \ac{CNN} accelerator designs to answer the major question of the paper whether SNNs surpass CNNs in terms of performance and energy efficiency when implemented on the same FPGA platform and using a configuration requiring approximately equal area and likewise for other metrics.
Specifically, the classification accuracy of the trained and quantized nets as well as when run on hardware is the same.
We use the Keras framework~\cite{keras} to model and train the networks employed for both accelerator types.
Likewise, we try to match the FPGA resource requirements of the designs in terms of LUTs, registers, \acp{BRAM}, and DSPs.
For synthesis settings, we use the Xilinx xc7z020-1clg400c part found on the PYNQ-Z1 board as well as the ZCU102 board with a larger FPGA chip (xczu9eg-ffvb1156-2-e) to evaluate the scalability of the approach.
The objectives we evaluate are execution time for classification, power, and, consequently, the energy required per sample.


First of all, we identify the corresponding configuration options to match FPGA resources.
For the SNN accelerator architecture proposed by~\citeauthor{sommer}~\cite{sommer}, this is the parallelization factor \(P\) as well as the \ac{AEQ} depth \(D\).
There is one \ac{AEQ} per \ac{PE}, which is replicated \(P\) times from ($P$ ranging from 1 to 16).
The depth \(D\) indicates that each \ac{AEQ} is sized to be able to hold \(D\) spike events.

For FINN, the effective settings are the SIMD values \(Q_l\) and the number of PEs per layer \(P_l\).

For the first experiment, we use the MNIST dataset for training and evaluation for both \acp{SNN} and corresponding traditional \acp{CNN} implementations. We chose this dataset as it is a commonly used benchmark set  in the literature.
The net we use for the MNIST classification also has the same architecture on the \ac{SNN} and \ac{CNN} accelerator.
The difference is that for \ac{SNN}, the model is translated via \texttt{snntoolbox} \cite{rueckauer} to a spiking net with m-TTFS encoding.
This incurs an accuracy loss, which is, however, small (0.4\%).
See Table~\ref{tbl:nn-models} for an overview of used model architectures.
Following the same notation, the used net is as follows: 32C3-32C3-P3-10C3-10.
As such, we have three pairs of \((Q_l, P_l)\) values with \(l=0,1,2\), which is how we denote the CNN configuration.
Without loss of generality and for simplicity, only the convolutional layers are numbered with \(l\) in the following discussion.

The \acl{SNN} accelerator by \citeauthor{sommer}~\cite{sommer} uses m-TTFS spike encoding and the IF neuron model with the constraint that neurons can only spike once and are not reset to zero afterward.
The number of algorithmic time steps is set to \(T = 4\) to achieve the noted accuracy.
The execution order of neurons within an \ac{SNN} inference can vary between accelerators.
The order is mathematically equivalent because inference works in a feed-forward manner in regular layers, including fully connected layers, convolutional layers, and max-pooling layers~\cite{syncnn}.
This equality holds only true as long as the IF neuron model is used.

In order to reduce the memory footprint, it is therefore viable to execute layer-by-layer, channel-by-channel in convolutional layers, and, finally, each layer for \(T\) repetitions.
In contrast, a parallelized implementation tends to be bottlenecked by available memory, which also affects energy requirements.
As such, using the IF neuron model, a neuron in, \eg a fully connected layer \(l\) has its membrane potential increased by a slope depending on the local weights and binary variables \(x_i^{(l-1)}\) at the previous layer (see Eq.~(\ref{eq:ifvm})).
Consider, for instance, layer \(l=1\) in the test net.
It can be run first by adding to the membrane potentials slopes computed from the spikes from layer \(l=0\), then doing the same again for three steps.

Also, note that none of the provided designs require any off-chip memory transfer of weights for comparability.
Only activations (MNIST sample images) are streamed into the architectures, and the classification result is taken via AXI interfaces.

Table~\ref{tbl:cnns} shows the considered FINN configurations with corresponding resource usage and accuracy.
The change in accuracy comes from different quantization settings during training resulting in a different weight bit width.
As can be seen, the bit width also has an effect on the number of resources needed for the \ac{MAC} units.
For instance, \cnnname{5} and \cnnname{6} differ only in bit width, and \cnnname{5} requires fewer LUTs and registers.

Table~\ref{tbl:snns} provides a set of synthesized SNN designs based on the SNN architecture by by \citeauthor{sommer}~\cite{sommer} that are comparable to the \ac{CNN} designs presented in Table~\ref{tbl:cnns}.
Both CNN and SNN designs are synthesized on the PYNQ-Z1.
The \acp{SNN} designs are characterized by the applied parallelization factor \(P\) as well as their memory configuration. For the above designs, only BRAMs are used as memories,, but we will show that using other means of storing  memory potentials and spikes can be beneficial in Section~\ref{sec:improvements}.
As can be seen, versions with 16-bit weights quickly become infeasible on the chosen target platform due to the excessive use of \acp{BRAM}.
In general, \acp{BRAM} can be identified as the resource which tends to be the limiting factor while only roughly half of the available LUT and register resources are used.

\pgfplotstableread[col sep=&, header=true]{./tables/cnnConfigs.tbl}{\cnnConfigTable}%
\pgfplotstableread[col sep=&, header=true]{./tables/snnConfigs.tbl}{\snnConfigTable}%
\pgfplotstableread[col sep=&, header=true]{./tables/origPower.tbl}{\origPowerTable}%
\pgfplotstableread[col sep=&, header=true]{./tables/origPowerManual.tbl}{\origPowerTableManual}%

\pgfplotstableset{
	columns/Design/.style={string type, column type={l}},
    columns/AccuracyMNIST/.style={postproc cell content/.style={@cell content=##1\%}, column name={Accuracy}},
    columns/AccuracySVHN/.style={postproc cell content/.style={@cell content=\ifstrequal{##1}{}{--}{##1\%}}, column name={Accuracy}},
    columns/AccuracyCIFAR/.style={postproc cell content/.style={@cell content=\ifstrequal{##1}{}{--}{##1\%}}, column name={Accuracy}},
    columns/FPS/W MNIST/.style={column name={FPS/W}},
    columns/FPS/W SVHN/.style={column name={FPS/W}},
    columns/FPS/W CIFAR/.style={column name={FPS/W}},
	f3/.style={fixed, precision=3},
	f2/.style={fixed, precision=2},
}

\begin{table}
  \caption{CNN configurations for the MNIST dataset generated with FINN for comparison with the SNN accelerator. The used platform is a PYNQ-Z1 board.}
\pgfplotstabletypeset[
]{\cnnConfigTable}
  \label{tbl:cnns}
\end{table}

\begin{table}
	\caption{SNN designs for the MNIST dataset analyzed within this work.
    The AEQ Depth is denoted by $D$, the degree of parallelism by $P$.
    The used platform is a PYNQ-Z1 board.}

	\pgfplotstabletypeset[
	columns/Parallelization/.style={column name={$P$}},
	columns/Depth/.style={column name={$D$}},
	every row 5 column 0/.style={
		postproc cell content/.style={@cell content=##1\textsuperscript{$\dagger$}}
	},
	]{\snnConfigTable}

  \label{tbl:snns}
\end{table}

\subsection{Evaluation of Latency and Power}

To determine the latency for sample classification, we run both FINN and \ac{SNN} accelerators in a simulator (Vivado).
FINN designs always require the same amount of cycles to complete, given the same streaming control signals, regardless of the input sample.
However, due to the nature of \acp{SNN}, latency cannot be measured as a single number in this case, as different samples generate different numbers of spikes.
Since sparse \ac{SNN} acceleration, put simply, processes spikes from queues until the queues are depleted, latency is highly dependent on data.
To measure this effect and enable a fair comparison of \ac{SNN} and \ac{CNN} approaches, we run the accelerator with 1,000 input images from the MNIST dataset {to get a good picture of the distribution of latencies depending on the input class/digit.
The results are visualized as a histogram in Figure~\ref{fig:origlatency}.
The bars represent the number of samples for which the latency (depicted on the y-axis) has been measured.
The red line is the latency of the corresponding FINN accelerator with similar resource usage.
As can be seen, the \snneightbram{} design is faster than \cnnname{4} for a majority of the input samples.
Frequencies have been set fixed to 100~MHz for both designs for comparability.
Maximum achievable frequencies vary from 120 to 130~MHz for \ac{SNN} and are about 105~MHz for the \ac{CNN} designs.

Figure~\ref{fig:latencyperclass} shows an evaluation of how different classes in the MNIST dataset affect the number of spikes generated per inference.
It can be seen that the class for the \texttt{1} digit is an outlier while the others are roughly equal.
This is due to the low number of pixels in the input feature map that are encoded to represent a spike before the SNN begins processing after thresholding.
Consequently, and depending on neuron/kernel weights, fewer spikes are also generated in subsequent layers and algorithmic time steps.
This shows that the execution time or energy consumption of an SNN is variable and depends on the input.

To determine the required electrical power of a design, we use the Vivado Power Estimator and focus on the dynamic power.
This tool allows for the use of post-implementation timing simulation data.
That is, the routed design is simulated using actual MNIST sample data, and the signal timings are recorded in a file.
This file can subsequently be input into the Power Estimator.
This is called vector-based estimation, while the purely statistical use of the Power Estimator results in a vector-less estimation.

As a result, here, just like with latency, the result depends on the input data, which is why we perform this estimation for multiple MNIST samples, both for CNN and SNN.
In the CNN case, we record power consumptions varying with less than 0.01W.
By contrast, the SNN accelerator does show significant variations depending on input data.
See Figure~\ref{fig:origpower} for a histogram visualization of the result.
Energy is determined by multiplying the execution time by the determined power.

See Table~\ref{tbl:origpower} for a detailed listing of the estimated power consumptions.
Dynamic power is further divided into power used for nets belonging to clocks, signals between slices as well as \acp{BRAM}.
Note that the \ac{BRAM} reading represents a very large portion of the total Watts reported.
In fact, \snneightbram{}, still better regarding latency in most cases, is worse than \cnnname{4} by a factor of about 4 regarding power consumption.
This is why we focus on analyzing and improving this metric.

\begin{figure}
	\begin{subfigure}{.32\linewidth}
		\begin{tikzpicture}
			\begin{axis}[
				ybar,
				width=\linewidth,
				xmin=0,
				ymin=0,
				ylabel={\#input samples},
				xlabel={Latency $\times$100}
				]
				\addplot+ [hist={bins=10}] table[y index=0] {data/hist_snn1.csv};
				\draw[dashed, red, ultra thick] (axis cs:515,0) -- 
					(axis cs:515,300);
			\end{axis}
		\end{tikzpicture}
        \caption{\snnonebram{} vs.\ \cnnname{2}}
	\end{subfigure}
	\begin{subfigure}{.32\linewidth}
		\begin{tikzpicture}
			\begin{axis}[
				ybar ,
				width=\linewidth,
				xmin=0,
				ymin=0,
				ylabel={\#input samples},
				xlabel={Latency $\times$100}
				]
				\addplot+ [hist={bins=10}] table[y index=0] {data/hist_latency-snn4-bram.csv};
                \draw[dashed, red, ultra thick] (axis cs:428,0) -- 
                	(axis cs:428,300);
			\end{axis}
		\end{tikzpicture}
        \caption{\snnfourbram{} vs. \cnnname{5}}
	\end{subfigure}
	\begin{subfigure}{.32\linewidth}
		\begin{tikzpicture}
			\begin{axis}[
				ybar ,
				width=\linewidth,
				xmin=0,
				ymin=0,
				ylabel={\#input samples},
				xlabel={Latency $\times$100}
				]
				\addplot+ [hist={bins=10}] table[y index=0] {data/hist_snn8.csv};
                \draw[dashed, red, ultra thick] (axis cs:387,0) -- 
                	(axis cs:387,300);
			\end{axis}
		\end{tikzpicture}
        \caption{\snneightbram{} vs. \cnnname{4}}
	\end{subfigure}
    \caption{Latency comparison of three SNN implementations (\snnonebram{}, \snnfourbram{}, and \snneightbram{}) and three CNN implementations of comparable resource usage (\cnnname{2}, \cnnname{5}, and \cnnname{4}). The histograms were generated measuring the latency for 1,000 images taken from the MNIST data set. The CNN implementations' latency does not depend on the input data and is visualized by the vertical dashed red line.}
    \label{fig:origlatency}
\end{figure}
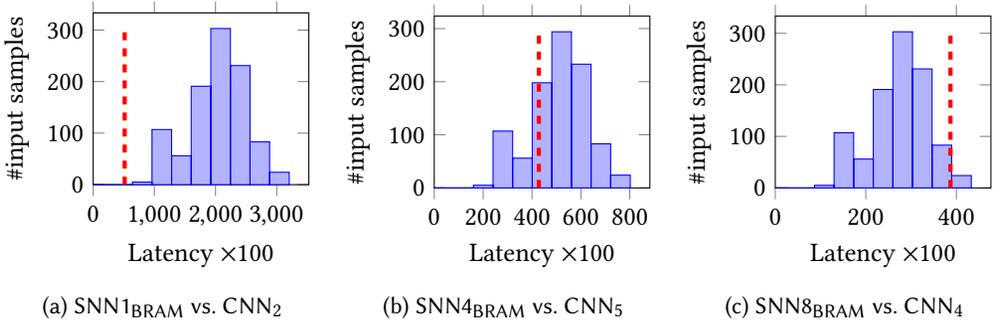

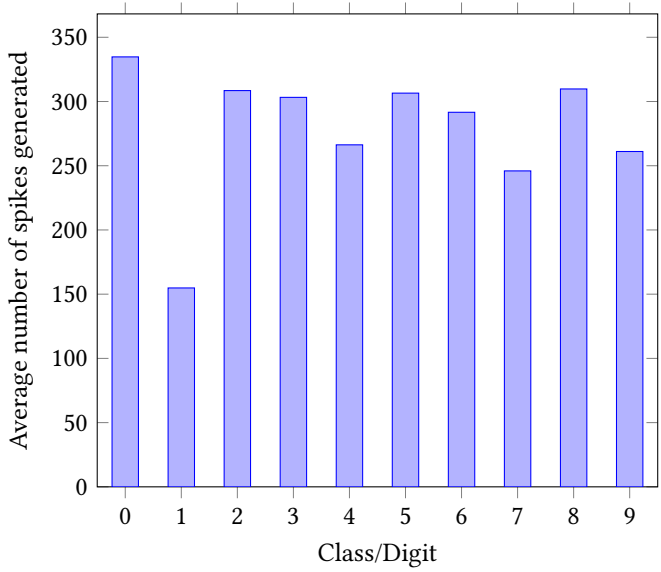
\begin{figure}
	\centering
	\begin{tikzpicture}
		\begin{axis}[
			xmin=-.5,
			xmax=9.5,
			ymin=0,
			ybar, 
			width=.65\linewidth,
			xtick=data,
			xlabel={Class/Digit},
			ylabel={Average number of spikes generated},
			]
			\addplot+ table[y index=3, fill] {data/perclass.csv};
		\end{axis}
	\end{tikzpicture}
  \caption{Average number of spikes generated during inference per class for the MNIST data set using \snneightbram.}
  \label{fig:latencyperclass}
\end{figure}

\begin{table}
	\caption{Vector-based estimation of the power of different designs.
    For SNNs, we report the minimum and maximum values.
    The actual distributions of these values are shown in Figure~\ref{fig:origpower}.}


	\pgfplotstabletypeset[
		string type,
	]{\origPowerTableManual}

\label{tbl:origpower}
\end{table}

\begin{figure}
	\begin{subfigure}{.45\linewidth}
		\centering
		\begin{tikzpicture}
	\begin{axis}[
		ybar ,
		width=\linewidth,
		xmin=0.10,
		ymin=0,
		ylabel={\#input samples},
		xlabel={Power [W]}
		]
		\addplot+ [hist={bins=60}] table[y index=0] {data/hist_power-snn4-bram.csv};
        \draw[dashed, red, ultra thick] (axis cs:0.107,0) -- coordinate[pos=.7, pin={[pin distance=13pt, black, pin edge={solid, black}, outer sep=0pt, inner sep=0pt]45:\cnnname{5}}] (snnComp) (axis cs:0.107,32);
	\end{axis}
\end{tikzpicture}
        \caption{Power \snnfourbram{} vs. \cnnname{5}}
	\end{subfigure}
    \begin{subfigure}{.45\linewidth}
		\centering
		\begin{tikzpicture}
	\begin{axis}[
		ybar ,
		width=\linewidth,
		xmin=0.04,
		ymin=0,
		ylabel={\#input samples},
		xlabel={Energy [mJ]}
		]
		\addplot+ [hist={bins=10}] table[y index=0] {data/hist_energy-snn4-bram.csv};
        \draw[dashed, red, ultra thick] (axis cs:0.0458,0) -- coordinate[pos=.7, pin={[pin distance=13pt, black, pin edge={solid, black}, outer sep=0pt, inner sep=0pt]45:\cnnname{5}}] (snnComp) (axis cs:0.0458,28);
	\end{axis}
\end{tikzpicture}
        \caption{Energy \snnfourbram{} vs. \cnnname{5}}
	\end{subfigure}

\vspace{2em}
	\begin{subfigure}{.45\linewidth}
		\centering
		\begin{tikzpicture}
	\begin{axis}[
		ybar ,
		width=\linewidth,
		xmin=0.10,
		ymin=0,
		ylabel={\#input samples},
		xlabel={Power [W]}
		]
		\addplot+ [hist={bins=60}] table[y index=0] {data/hist_power-snn8-bram.csv};
        \draw[dashed, red, ultra thick] (axis cs:0.119,0) -- coordinate[pos=.7, pin={[pin distance=13pt, black, pin edge={solid, black}, outer sep=0pt, inner sep=0pt]45:\cnnname{4}}] (snnComp) (axis cs:0.119,40);
	\end{axis}
\end{tikzpicture}
        \caption{Power \snneightbram{} vs. \cnnname{4}}
	\end{subfigure}
    \begin{subfigure}{.45\linewidth}
		\centering
		\begin{tikzpicture}
	\begin{axis}[
		ybar ,
		width=\linewidth,
		xmin=0.04,
		ymin=0,
		ylabel={\#input samples},
		xlabel={Energy [mJ]}
		]
		\addplot+ [hist={bins=10}] table[y index=0] {data/hist_energy-snn8-bram.csv};
        \draw[dashed, red, ultra thick] (axis cs:0.0461,0) -- coordinate[pos=.7, pin={[pin distance=13pt, black, pin edge={solid, black}, outer sep=0pt, inner sep=0pt]45:\cnnname{4}}] (snnComp) (axis cs:0.0461,40);
	\end{axis}
\end{tikzpicture}
        \caption{Energy \snneightbram{} vs. \cnnname{4}}
	\end{subfigure}
    \caption{Comparison of power and energy between the \snnfourbram{} and \cnnname{5} as well as the \snneightbram{} and \cnnname{4} accelerators. The red line shows the power or energy per classification of the CNN accelerator, while the SNN data is plotted as a histogram over multiple MNIST samples since it is dependent on input data.}
    \label{fig:origpower}
\end{figure}
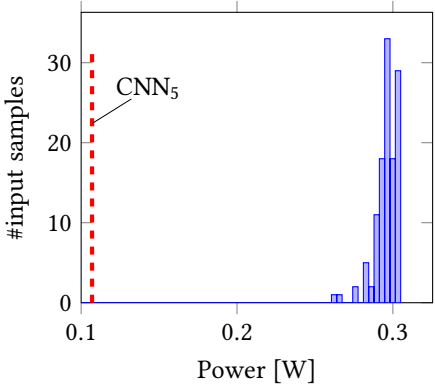
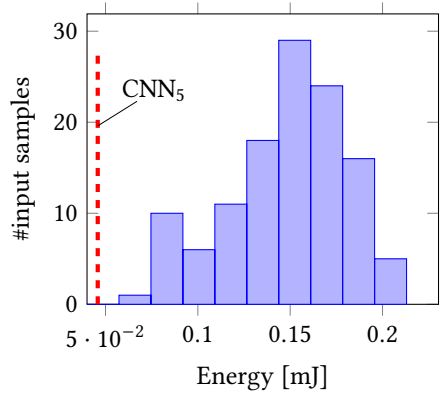
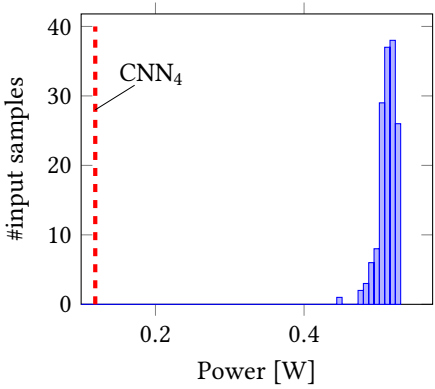
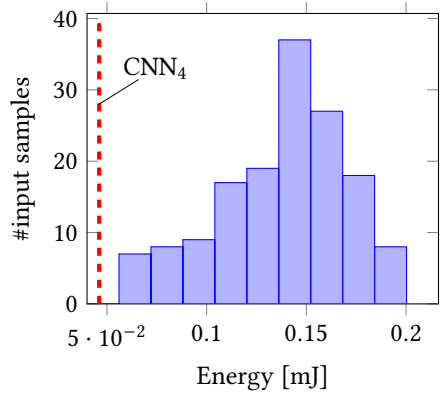

\subsection{Computation of \ac{BRAM} Usage}
FINN designs require much fewer \acp{BRAM} than the SNN implementations because neurons are only stored as intermediate results and not as a matrix of membrane potentials.
Also, the \acp{AEQ} take up roughly half the \ac{BRAM} resources as well.
Both need to be replicated \(P\)~times to increase throughput and are not filled 100\%.

Xilinx \ac{BRAM} primitives have a fixed size but can be used to store words of differing lengths. The number of words, depending on the word width $w$, in a \ac{BRAM} is computed as
\begin{equation}
	\mathrm{\#words}(w) = \begin{cases} 1024 & \text{if } 18 < w \leq 36  \\
		2048 & \text{if } 9 < w \leq 18 \\
        4096 & \text{if } 4 < w \le 8 \\
        8192 & \text{if } 2 < w \le 4 \\
        16384 & \text{if } w = 2 \\
        32768 & \text{if } w = 1 \end{cases}.
     \label{eq:bramwords}
\end{equation}

The smallest unit possible for instantiating \acp{BRAM} is half a \ac{BRAM}.
Hence, the number of \acp{BRAM} required to store $n$ words is
\begin{equation}
	\left\lceil n\right\rceil_{\mathrm{BRAM}} = \frac{\lceil2\cdot n\rceil}{2}.
\end{equation}

These numbers allow deriving the number of required \acp{BRAM} for a given kernel size $K$, the degree of parallelization $P$, queue depth $D$, and word width $w$ to be
\begin{align}
	\#\mathrm{BRAM} &= P\cdot K \cdot \left\lceil\frac{D}{\mathrm{\#words}(w)}\right\rceil_{\mathrm{BRAM}}. \label{eq:numBram}
\end{align}

This value can directly be used to determine the number of \acp{BRAM} for the Address Event Queues, i.e., we have $\#\mathrm{BRAM}_{\mathrm{AEQ}}=\#\mathrm{BRAM}$. 
Here, the word width is the number of bits required to store one spike event, \ie \(w_\mathrm{AE}\).
As the computation of the membrane potentials involves values pre- and post-computation, the number of required \acp{BRAM} is doubled, \ie $\#\mathrm{BRAM}_{\mathrm{Membrane}}=2\cdot \#\mathrm{BRAM}$

\begin{table}
		\caption{BRAM usage for different SNN designs.}
	\label{tbl:bramsorig}
	\begin{tabular}{lrrrrrrr}
		\toprule
		Name 	& \(D\) & $D_{\mathrm{Membrane}}$ 	& \(w\) & $w_{\mathrm{Membrane}}$ 	& \(P\)	& $\#\mathrm{BRAM}_{\mathrm{AEQ}}$ & $\#\mathrm{BRAM}_{\mathrm{Membrane}}$ \\ \midrule
        \snnonebram{} (\(w=16\)) & 6100 & 256 & 10 & 16 & 1 & 27 & 9  \\
        \snnfourbram{} & 2048 & 256 & 10 & 8& 4 & 36 & 36\\
        \snneightbram{} & 750 & 256 & 10 & 8& 8 & 36 & 72\\\bottomrule
	\end{tabular}

\end{table}

Additionally, kernel and fully connected layer weights must be stored.
However, these memories are read-only and subject to optimizations by the synthesis tool.
It turns out that 2.5 BRAM primitives can fit all weights per PE.
Therefore, for configurations where it is feasible to do so, a maximum of \(2.5\cdot P\) BRAMs is added to the total number.

%

\section{Architectural Improvements}
\label{sec:improvements}

In this section, we present extensions and improvement of the SNN accelerator designs as analyzed in Section~\ref{sec:experiments}. Additionally,  the study is extended to include both \acp{SNN} and \acp{CNN} accelerators performing classification also for more complex networks, i.e., the SVHN and CIFAR-10 datasets.
Since SVHN and CIFAR-10 are more difficult tasks than MNIST, larger models are used for these and implemented using both FINN and as SNNs.
Refer to Table~\ref{tbl:nn-models} for an overview of the datasets and models used for each.
The size of each model is measured in the number of weight/bias parameters output by Keras.

The architectures are chosen as a trade-off between size and classification accuracy in each case to provide the opportunity to test the scalability of the implementation approaches.
Due to comparability, one of our target platforms is the  PYNQ board focused on edge applications and providing a small FPGA (xc7z020-1clg400c).
For the reason of resource scarcity, well-known NN models such as VGG or LeNet are difficult to implement.
For example, a VGG-5 implementation has a total of 2,707,882 parameters, and is not implementable because our CIFAR-10 model already leads to maximum resource usage for BRAMs for larger parallelization factors.
However, the chosen models are based roughly on the LeNet architecture.

In order to study also larger networks, we provide a second suite of experiments based on the ZCU102 Zynq UltraScale+ board providing a xczu9eg-ffvb1156-2-e chip.

\begin{table}
	\caption{Overview of model architectures used for datasets MNIST, SVHN, and CIFAR-10.
    Here, \(n\mathrm{C}k\) denotes a convolutional layer with \(n\) kernels of size \(k\times k\), \(\mathrm{P}n\) a pooling layer with a window size of \(n\), and just \(n\) a fully connected layer with \(n\) neurons.
        The last two columns show Keras's classification accuracy, including quantization effects before and after conversion using the \texttt{snntoolbox} \cite{rueckauer}.}
    \resizebox{\columnwidth}{!}{%
\begin{tabular}{lcccc}
	\toprule
    &&& \multicolumn{2}{c}{\tblheader{Accuracy}} \\
    \cmidrule{4-5}
    \tblheader{Dataset} & \tblheader{Model Architecture} & \tblheader{Num. Params} & \tblheader{Keras} & \tblheader{snntoolbox} \\
    \midrule
    MNIST & 32C3-32C3-P3-10C3-10 & 20,568 & 97.8\% & 98.2\% \\
    SVHN & 1C3-32C3-32C3-P3-64C3-64C3-P3-128C3-128C3-10 & 297,966 & 91.7\% & 72.1\% \\
    CIFAR-10 & 32C3-32C3-P3-64C3-64C3-P3-128C3-128C3-128C3-10 & 446,122 & 80.1\% & 60.2\% \\
    \bottomrule
\end{tabular}}%
\label{tbl:nn-models}
\end{table}

\subsection{FPGA Memory Scalability Study}
Memory usage for \acp{SNN} can be divided into (a)~membrane potentials, (b)~data structures for storing spike sequences, and (c)~the read-only kernel and dense layer weights.
First of all, there is the option of synthesizing large memories as BRAMs or LUTRAMs, depending on the granularity and synthesis settings of the FPGA toolchain.
For Xilinx devices, BRAMs store 36K bits and can be configured to be accessed using 36-, 18-, 9-, 4-, 2-, or 1-bit words.
Also, it is possible to use halves of BRAMs, storing 18K bits.
Next, only one word can be read or written during one clock cycle.
If parallelized memory accesses are desired, BRAMs must be split for the sake of latency reduction, even though they might be sparsely occupied as a result.
For this reason, the number of BRAMs is determined not only by the amount of data to store but also by the parallelism in access patterns.
On the other hand, LUTRAMs can be instantiated in a much more fine-grained manner but not as energy-efficient when fully utilized compared to BRAMs.

Both RAM types require a substantial amount of power to drive.
Where is the point when it becomes more efficient to opt for LUTRAM rather than BRAM?
As an experiment, we created a BRAM test design, visualized in Figure~\ref{fig:bramtest}, which uses an array of \(R\) BRAM-based memories to store 8192 words of bit width \(w\). The output of the individual BRAMs is XOR'ed to compute an output word of width $w$ without incurring a measurable impact on energy.


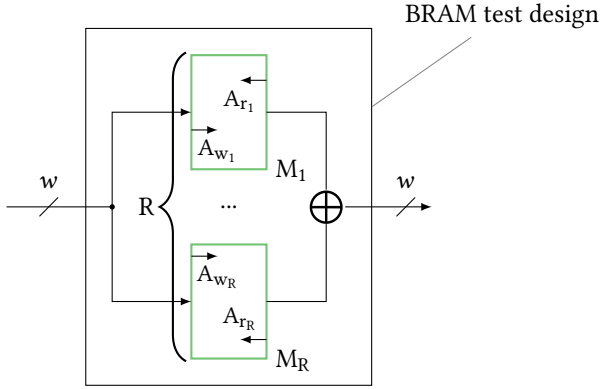
\begin{figure}
	\begin{tikzpicture}[
			bram/.style={
				draw=Greens-G,
				minimum height=1.5cm, 
				thick, 
				minimum width=1cm, 
				outer sep=0pt,
				rectangle},
			dot/.style={fill,inner sep=0pt, outer sep=0pt,  minimum width=.2em, circle,}
		]
		\node[bram, label=-45:M\textsubscript{1}] (bram0) {};
		\node[below=1em of bram0] (dots) {...};
		\node[bram, below=1em of dots, label=-45:M\textsubscript{R}] (bramR) {};

		\node[right=3em of dots, anchor=center, inner sep=0pt, outer sep=0pt] (plus) {\Huge$\oplus$};
		
		\draw (bram0) -| (plus);
		\draw (bramR) -| (plus);
		
		\coordinate[xshift=-3em] (helper) at (bram0.west);
		
		\draw[latex-latex] (bram0) -- (helper) -- coordinate[pos=.5] (inH) (helper |- bramR) -- (bramR);
		
		\draw (inH) -- node[dot, pos=0] {} node[pos=.6, draw, strike out, label=90:$w$] {} ++(-4em,0);
		\draw[-latex] (plus) -- node[pos=.7, draw, strike out, label=90:$w$] {} ++(4em,0);
		
		\draw[decorate, thick, decoration={brace, aspect=.5, amplitude=10pt}] ($(bramR.south west)+(-.2em,-.1em)$) -- node[pos=.5,xshift=-15pt] {R} ($(bram0.north west)+(-.2em,.1em)$);
		
		\node[fit=(helper) (plus) (bramR) (bram0), draw, inner sep=1em, pin=45:{BRAM test design}] (box) {};

		\draw[-latex] (bram0.40) -- node[pos=1, below] {\small A\textsubscript{r\textsubscript{1}}}++(-1em,0);
		\draw[-latex] (bram0.205) -- node[pos=1, below] {\small A\textsubscript{w\textsubscript{1}}}++(1em,0);

		
		\draw[-latex] (bramR.-45) -- node[pos=1, above] {\small A\textsubscript{r\textsubscript{R}}}++(-1em,0);
		\draw[-latex] (bramR.130) -- node[pos=1, below] {\small A\textsubscript{w\textsubscript{R}}}++(1em,0);
	\end{tikzpicture}
	\caption{BRAM test design: An array of $R$ memory blocks M\textsubscript{i} is employed with energy measurements. Each memory block might be composed of several BRAMs to store a total of $D$ words. The design allows for the constant writing of one value using the write pointers A\textsubscript{w\textsubscript{i}} and reading of individual values using the read pointers A\textsubscript{r\textsubscript{i}}. The individual output words are XOR'ed to obtain a single output of width $w$.}
	\label{fig:bramtest}
\end{figure}

Multiple access patterns are possible: The memories can be written with the incoming data from the streaming interface with bit width \(w\) or they can be set to be read-only.
In both cases, they are pre-initialized with random data.
The read pointers and write pointers A\textsubscript{r\textsubscript{i}}  and A\textsubscript{w\textsubscript{i}}, respectively, are initialized to different positions.
Here, all memories are written simultaneously in a single clock cycle with the same input word.

For the conducted experiments, the setting was to continuously read from all memories, \ie in every clock cycle.
We synthesized variants using (a) actual BRAMs or (b) LUTRAMs to investigate when to choose which type of memory. We varied the bit-width $w$ from 1 to 36 and measured the power.
The resulting power measurements are depicted in Figure~\ref{fig:bramlutramcomp}.
As can be seen, LUTRAMs scale linearly with the bit width \(w\) while BRAMs tend to effect an increase in power whenever the bit thresholds given in Eq.~(\ref{eq:bramwords}) are reached.
Note that words with a width of 10 bits can, for instance, also be synthesized to be composed of 2 words with a width of 3 each and 1 word with a width of 4, resulting in 3 BRAMs with a more favorable configuration than a single BRAM storing 10-bit words.

A major factor in deciding whether to use LUTRAMs or BRAMs is the depth \(D\) of each memory row.
As can be seen in Figure~\ref{fig:bramlutramcomp}, LUTRAMs perform better than BRAMs whenever words do not fit exactly into the available aspect ratios of BRAMs.
For instance, \(D = 256\) is not favorable for BRAMs as it leads to multiple half BRAMs being synthesized, which are not fully used.

In the following, we use both of these insights to improve the memory architecture of the examined SNN accelerator: Reduce inefficient BRAM usage for small depths and drop word lengths below the aspect ratio thresholds.

\begin{figure}
	\begin{subfigure}{.9\linewidth}
		\centering
	\scalebox{.8}{%
		\begin{tikzpicture}
	\begin{axis}[
		ybar ,
		width=\linewidth,
		xmin=0.04,
		ymin=0,
yticklabel style={
	/pgf/number format/fixed,
	/pgf/number format/precision=2
},
scaled y ticks=false,
        ylabel={Power [W]},
		xlabel={Bit Width $w$},
		cycle list/Set1-4,
		legend cell align={left},
  			bram/.style={mark=square*, only marks,
			mark options={
				scale=1.3,
			}
		},
		lut/.style={mark=triangle*, only marks,
			mark options={
				scale=1.8,
			}
		}
		]

		\addplot+ [bram] table {data/bramlutram8192-1-False.csv};
        \addlegendentry{BRAM, R=1}
		\addplot+ [bram] table {data/bramlutram8192-2-False.csv};
        \addlegendentry{BRAM, R=2}
		\addplot+ [bram] table {data/bramlutram8192-8-False.csv};
        \addlegendentry{BRAM, R=8}
		\addplot+ [bram] table {data/bramlutram8192-16-False.csv};
        \addlegendentry{BRAM, R=16}
   		\addplot+ [lut] table {data/bramlutram8192-1-True.csv};
        \addlegendentry{LUT, R=1}
        \addplot+ [lut] table {data/bramlutram8192-2-True.csv};
        \addlegendentry{LUT, R=2}
        \addplot+ [lut] table {data/bramlutram8192-8-True.csv};
        \addlegendentry{LUT, R=8}
        \addplot+ [lut] table {data/bramlutram8192-16-True.csv};
        \addlegendentry{LUT, R=16}
	\end{axis}
\end{tikzpicture}
}
        \caption{ \(D = 8192\)}
	\end{subfigure}

\vspace{2em}
    \begin{subfigure}{.9\linewidth}
      \centering
  	\scalebox{.8}{
  		\begin{tikzpicture}
  		\begin{axis}[
  			ybar ,
  			xmin=0.04,
			width=\linewidth,
  			ymin=0,
  			yticklabel style={
  				/pgf/number format/fixed,
  				/pgf/number format/precision=2
  			},
  			scaled y ticks=false,
  			ylabel={Power [W]},
  			xlabel={Bit Width $w$},
  			cycle list/Set1-4,
  			legend cell align={left},
  			bram/.style={mark=square*, only marks,
				mark options={
  					scale=1.3,
  				}
  			},
  			lut/.style={mark=triangle*, only marks,
				mark options={
  					scale=1.8,
  				}
  			}
  			]
  			
  			\addplot+ [bram] table {data/bramlutram256-1-False.csv};
  			\addlegendentry{BRAM, R=1}
  			\addplot+ [bram] table {data/bramlutram256-2-False.csv};
  			\addlegendentry{BRAM, R=2}
  			\addplot+ [bram] table {data/bramlutram256-8-False.csv};
  			\addlegendentry{BRAM, R=8}
  			\addplot+ [bram] table {data/bramlutram256-16-False.csv};
  			\addlegendentry{BRAM, R=16}
  			\addplot+ [lut] table {data/bramlutram256-1-True.csv};
  			\addlegendentry{LUT, R=1}
  			\addplot+ [lut] table {data/bramlutram256-2-True.csv};
  			\addlegendentry{LUT, R=2}
  			\addplot+ [lut] table {data/bramlutram256-8-True.csv};
  			\addlegendentry{LUT, R=8}
  			\addplot+ [lut] table {data/bramlutram256-16-True.csv};
  			\addlegendentry{LUT, R=16}
  		\end{axis}
  	\end{tikzpicture}
  }
        \caption{\(D = 256\)}
    \end{subfigure}

	\caption{Results of BRAM vs. LUTRAM power comparison for (a) $D=8192$ and (b) $D=256$.}
	\label{fig:bramlutramcomp}
\end{figure}
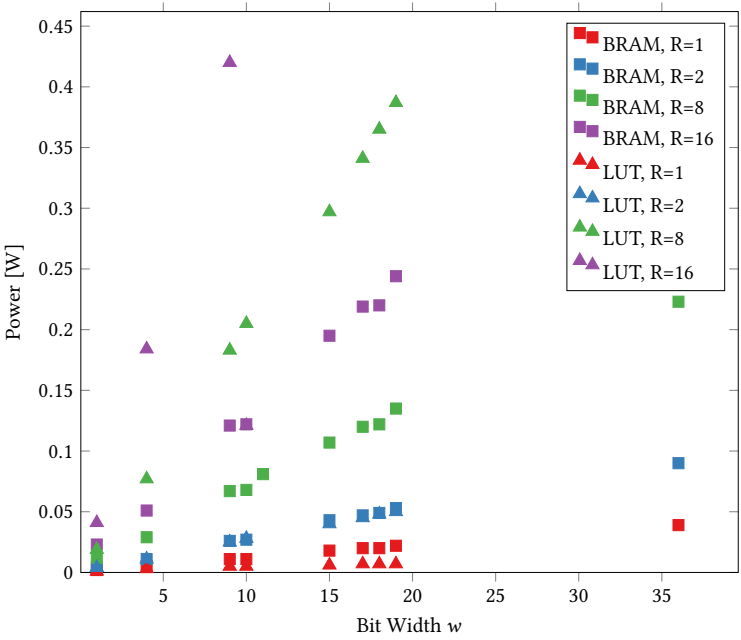
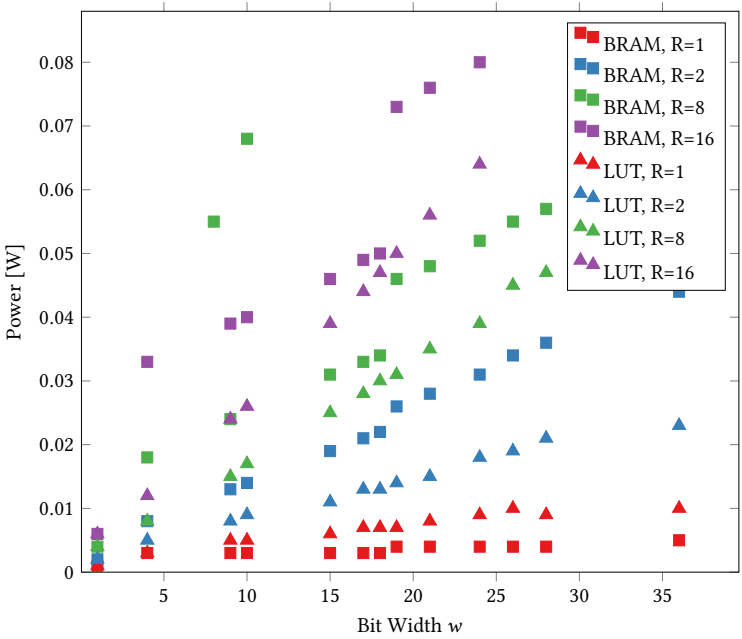

\subsection{Evaluation of Optimization Techniques}
When accelerating \acp{SNN} on FPGAs, we identify the membrane potentials as a source of inefficiency.
Unlike in \acp{CNN}, where all neurons are computed sequentially by way of performing matrix multiplications, in \acp{SNN}, all neuron potentials must be held in memory.
However, due to the high degree of parallelization, these are distributed across many memories.
For instance, we determine the number of words of membrane potential memory never to exceed 256 in our experiments.
Since BRAMs can hold 4096 8-bit words, this means an actual usage of only 6.25\%, which is very wasteful.
By changing the memory interlacing scheme to implement required memory blocks with low usage as LUTRAMs, the energy efficiency can be improved.
Refer to Table~\ref{tbl:improvements}, \eg the change between the original \snneightbram{} and the improved \snneightlutram{} design.
As can be seen, power can be reduced by about 15\%.
A side effect is the shift of resource usage from BRAMs to LUTs.
This creates an even more balanced design.

\begin{table}
	\caption{Resource usage and vector-less power estimation of base designs and improved designs.}
	\label{tbl:improvements}
\begin{tabular}{lNNNrrrrr}
	\toprule
  & & & & \multicolumn{5}{c}{\tblheader{Power [W]}} \\ \cmidrule{5-9}
  \tblheader{Design} & \tblheader{LUTs} & \tblheader{Regs.} & \tblheader{BRAMs} & \tblheader{Signals} & \tblheader{BRAMs} & \tblheader{Logic} & \tblheader{Clocks} & \tblheader{Total} \\ \midrule
  \cnnname{4}           & 20368 & 26886 & 14.5 & 0.039 & 0.012 & 0.036 & 0.035 & \textbf{0.122} \\
  \cnnname{5}           & 16793 & 17810 & 11 & 0.035 & 0.012 & 0.028 & 0.032 & \textbf{0.107} \\
  \snnfourbram{} &  4967 & 5019 & 76 & 0.041 & 0.185 & 0.027 & 0.030 & \textbf{0.283} \\
  \snnfourlutram{} &  9256 & 5669 & 40 & 0.068 & 0.099 & 0.041 & 0.034 & \textbf{0.242} \\
  \snnfourpacking{}  & 9436 & 5669 & 22 & 0.068 & 0.056 & 0.043 & 0.033 & \textbf{0.200} \\
  \snneightbram{} & 9649 & 9738 & 116 & 0.089 & 0.277 & 0.059 & 0.055 & \textbf{0.480} \\
  \snneightlutram{} & 18311 & 11080 & 44 & 0.146 & 0.106 & 0.091 & 0.062 & \textbf{0.405} \\
  \snneightpacking{} & 18311 & 11080 & 44 & 0.146 & 0.106 & 0.091 & 0.062 & \textbf{0.405} \\
  \bottomrule
\end{tabular}%
\label{tbl:improvements}
\end{table}

A major cause for wasted memory is the gap between the word sizes of Xilinx BRAM primitives.
This is most pronounced in the \ac{AEQ} implementations.
In Table~\ref{tbl:snns}, a word width of 10 bits causes each BRAM to hold only 2048 words, whereas it can hold 4096 9-bit words.
This is an issue that can be overcome by reducing the word width by compressing spike events.

Therefore, we propose the use of an improved encoding of spikes as compressed coordinates \((i_c, j_c)\).
In the original work \cite{sommer}, two additional status bits were used to signify the segmentation of the \acp{AEQ}.
These can be done away with when recognizing that for a feature map of \(28\times 28\), since it is divided into windows of \(3\times 3\) due to the kernel size \(K=3\) in this case, actual coordinates can be encoded as the \emph{explicit} number as well as the \emph{implicit} window position given by the queue data structure the event is stored in.
Let \(W = 28\) be the feature map width.
For quadratic sizes, the required bit width for \(i_c\) is
\begin{equation}
\left\lceil\log_2\frac{W}{K}\right\rceil = 4.
\end{equation}
There exist 6 unused bit-patterns for both \(i_c\) and \(j_c\). These can be used to encode status information with minimal logic overhead.

\begin{figure}
	\begin{subfigure}{.45\linewidth}
		\centering
		\begin{tikzpicture}
	\begin{axis}[
		ybar ,
		width=\linewidth,
        scaled x ticks=false,
		xmin=0.10,
		ymin=0,
		ylabel={\#input samples},
        xlabel={Power [W]}
		]
		\addplot+ [hist={bins=10}] table[y index=0] {data/hist_power-snn4-packing.csv};
        \draw[dashed, red, ultra thick] (axis cs:0.107,0) -- coordinate[pos=.7, pin={[pin distance=13pt, black, pin edge={solid, black}, outer sep=0pt, inner sep=0pt]80:\cnnname{5}}] (snnComp) (axis cs:0.107,40);
	\end{axis}
\end{tikzpicture}
        \caption{Power \snnfourpacking{} vs. \cnnname{5}}
	\end{subfigure}
    \begin{subfigure}{.45\linewidth}
		\centering
		\begin{tikzpicture}
	\begin{axis}[
		ybar ,
		width=\linewidth,
        scaled x ticks=false,
		xmin=0.10,
		ymin=0,
		ylabel={\#input samples},
        xlabel={Power [W]}
		]
		\addplot+ [hist={bins=10}] table[y index=0] {data/hist_power-snn8-packing.csv};
        \draw[dashed, red, ultra thick] (axis cs:0.119,0) -- coordinate[pos=.7, pin={[pin distance=13pt, black, pin edge={solid, black}, outer sep=0pt, inner sep=0pt]45:\cnnname{4}}] (snnComp) (axis cs:0.119,90);
	\end{axis}
\end{tikzpicture}
        \caption{Power \snneightpacking{} vs. \cnnname{4}}
	\end{subfigure}

	\vspace{1.5em}
	\begin{subfigure}{.45\linewidth}
		\centering
		\begin{tikzpicture}
	\begin{axis}[
		ybar ,
		width=\linewidth,
        scaled x ticks=false,
		xmin=0.00,
		ymin=0,
		ylabel={\#input samples},
		xlabel={Energy [mJ]}
		]
		\addplot+ [hist={bins=6}] table[y index=0] {data/hist_energy-snn4-packing.csv};
        \draw[dashed, red, ultra thick] (axis cs:0.0458,0) -- coordinate[pos=.7, pin={[pin distance=13pt, black, pin edge={solid, black}, outer sep=0pt, inner sep=0pt]85:\cnnname{5}}] (snnComp) (axis cs:0.0458,42);
	\end{axis}
\end{tikzpicture}
        \caption{Energy \snnfourpacking{} vs. \cnnname{5}}
	\end{subfigure}
	\begin{subfigure}{.45\linewidth}
	\centering
	\begin{tikzpicture}
		\begin{axis}[
			ybar ,
			width=\linewidth,
			scaled x ticks=false,
			xmin=0.00,
			ymin=0,
			ylabel={\#input samples},
			xlabel={Energy [mJ]}
			]
			\addplot+ [hist={bins=6}] table[y index=0] {data/hist_energy-snn8-packing.csv};
			\draw[dashed, red, ultra thick] (axis cs:0.045,0) -- coordinate[pos=.7, pin={[pin distance=13pt, black, pin edge={solid, black}, outer sep=0pt, inner sep=0pt]84:{\cnnname{4}}}] (snnComp) (axis cs:0.045,75);
		\end{axis}
	\end{tikzpicture}
	\caption{Energy \snneightpacking{} vs. \cnnname{4}}
\end{subfigure}

	\vspace{1.5em}
    \begin{subfigure}{.45\linewidth}
	\centering
	\begin{tikzpicture}
		\begin{axis}[
			ybar ,
			width=\linewidth,
			scaled x ticks=false,
			xmin=0.04,
			ymin=0,
			ylabel={\#input samples},
			xlabel={FPS/W}
			]
			\addplot+ [hist={bins=10}] table[y index=0] {data/hist_fpsw-snn4-packing.csv};
			\draw[dashed, red, ultra thick] (axis cs:21809,0) -- coordinate[pos=.7, pin={[pin distance=13pt, black, pin edge={solid, black}, outer sep=0pt, inner sep=0pt]99:\cnnname{5}}] (snnComp) (axis cs:21809,40);
		\end{axis}
	\end{tikzpicture}
	\caption{FPS/W \snnfourpacking{} vs. \cnnname{5}}
\end{subfigure}
    \begin{subfigure}{.45\linewidth}
		\centering
		\begin{tikzpicture}
	\begin{axis}[
		ybar ,
		width=\linewidth,
        scaled x ticks=false,
		xmin=0.04,
		ymin=0,
		ylabel={\#input samples},
		xlabel={FPS/W}
		]
		\addplot+ [hist={bins=10}] table[y index=0] {data/hist_fpsw-snn8-packing.csv};
        \draw[dashed, red, ultra thick] (axis cs:22218,0) -- coordinate[pos=.7, pin={[pin distance=13pt, black, pin edge={solid, black}, outer sep=0pt, inner sep=0pt]99:\cnnname{4}}] (snnComp) (axis cs:22218,75);
	\end{axis}
\end{tikzpicture}
        \caption{FPS/W \snneightpacking{} vs. \cnnname{4}}
	\end{subfigure}

    \caption{Comparison of energy and FPS/W between the \snnfourpacking{} and \snneightpacking{} and their corresponding CNN designs (\cnnname{5} and \cnnname{4}, respectively).}
    \label{fig:improvedpower}
\end{figure}
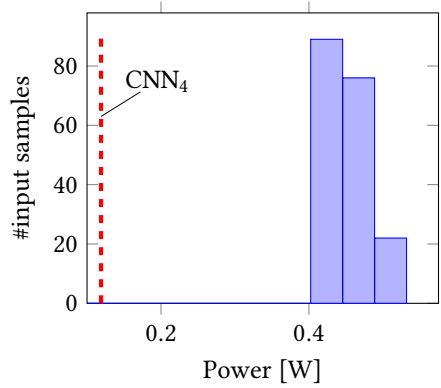
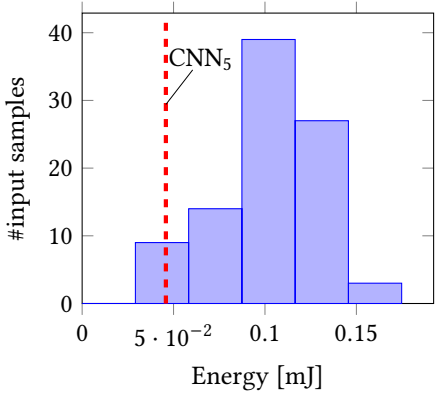
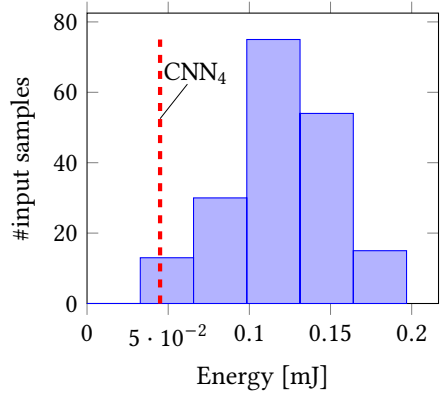
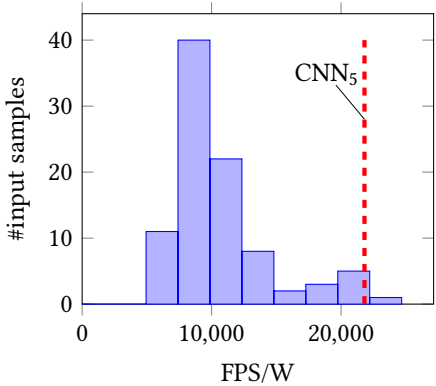
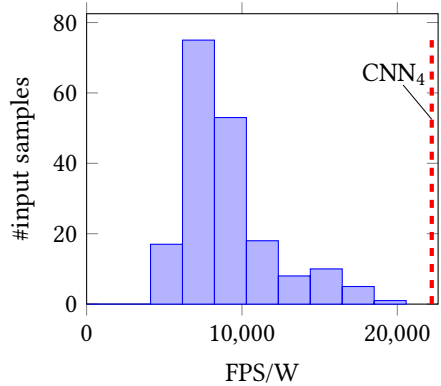

	\begin{figure}
		\begin{subfigure}{.45\linewidth}
			\centering
			\begin{tikzpicture}
				\begin{axis}[
					ybar ,
					width=\linewidth,
					scaled x ticks=false,
					xmin=0.10,
					ymin=0,
					xmax=0.52,
					ylabel={\#input samples},
					xlabel={Power [W]}
					]
					\addplot+ [hist={bins=10}] table[y index=0] {data/hist_power-snn4-svhn.csv};
					\draw[dashed, red, ultra thick] (axis cs:0.450,0) -- coordinate[pos=.7, pin={[pin distance=13pt, black, pin edge={solid, black}, outer sep=0pt, inner sep=0pt]100:\cnnname{7}}] (snnComp) (axis cs:0.450,50);
				\end{axis}
			\end{tikzpicture}
			\caption{Power \snnfoursvhn{} vs. \cnnname{7}}
		\end{subfigure}
		\begin{subfigure}{.45\linewidth}
			\centering
			\begin{tikzpicture}
				\begin{axis}[
					ybar ,
					width=\linewidth,
					scaled x ticks=false,
					xmin=0.10,
					ymin=0,
					xmax=0.7,
					ylabel={\#input samples},
					xlabel={Power [W]}
					]
					\addplot+ [hist={bins=10}] table[y index=0] {data/hist_power-snn8-svhn.csv};
					\draw[dashed, red, ultra thick] (axis cs:0.535,0) -- coordinate[pos=.6, pin={[pin distance=13pt, black, pin edge={solid, black}, outer sep=0pt, inner sep=0pt]110:\cnnname{8}}] (snnComp) (axis cs:0.535,50);
				\end{axis}
			\end{tikzpicture}
			\caption{Power \snneightsvhn{} vs. \cnnname{8}}
		\end{subfigure}
		
		\vspace{1.5em}
		\begin{subfigure}{.45\linewidth}
			\centering
			\begin{tikzpicture}
				\begin{axis}[
					ybar ,
					width=\linewidth,
					scaled x ticks=false,
					xmin=0.00,
					ymin=0,
					ylabel={\#input samples},
					xlabel={Energy [mJ]}
					]
					\addplot+ [hist={bins=6}] table[y index=0] {data/hist_energy-snn4-svhn.csv};
					\draw[dashed, red, ultra thick] (axis cs:2.16,0) -- coordinate[pos=.6, pin={[pin distance=13pt, black, pin edge={solid, black}, outer sep=0pt, inner sep=0pt]26:\cnnname{7}}] (snnComp) (axis cs:2.16,37);
				\end{axis}
			\end{tikzpicture}
			\caption{Energy \snnfoursvhn{} vs. \cnnname{7}}
		\end{subfigure}
		\begin{subfigure}{.45\linewidth}
			\centering
			\begin{tikzpicture}
				\begin{axis}[
					ybar ,
					width=\linewidth,
					scaled x ticks=false,
					xmin=0.00,
					ymin=0,
					ylabel={\#input samples},
					xlabel={Energy [mJ]}
					]
					\addplot+ [hist={bins=6}] table[y index=0] {data/hist_energy-snn8-svhn.csv};
					\draw[dashed, red, ultra thick] (axis cs:1.87,0) -- coordinate[pos=.7, pin={[pin distance=13pt, black, pin edge={solid, black}, outer sep=0pt, inner sep=0pt]25:{\cnnname{8}}}] (snnComp) (axis cs:1.87,30);
				\end{axis}
			\end{tikzpicture}
			\caption{Energy \snneightsvhn{} vs. \cnnname{8}}
		\end{subfigure}
		
		\vspace{1.5em}    
		\begin{subfigure}{.45\linewidth}
			\centering
			\begin{tikzpicture}
				\begin{axis}[
					ybar ,
					width=\linewidth,
					scaled x ticks=false,
					xmin=0.04,
					ymin=0,
					ylabel={\#input samples},
					xlabel={FPS/W}
					]
					\addplot+ [hist={bins=10}] table[y index=0] {data/hist_fpsw-snn4-svhn.csv};
					\draw[dashed, red, ultra thick] (axis cs:463,0) -- coordinate[pos=.7, pin={[pin distance=13pt, black, pin edge={solid, black}, outer sep=0pt, inner sep=0pt]135:\cnnname{7}}] (snnComp) (axis cs:463,30);
				\end{axis}
			\end{tikzpicture}
			\caption{FPS/W \snnfoursvhn{} vs. \cnnname{7}}
		\end{subfigure}
		\begin{subfigure}{.45\linewidth}
			\centering
			\begin{tikzpicture}
				\begin{axis}[
					ybar ,
					width=\linewidth,
					scaled x ticks=false,
					xmin=0.04,
					ymin=0,
					ylabel={\#input samples},
					xlabel={FPS/W}
					]
					\addplot+ [hist={bins=10}] table[y index=0] {data/hist_fpsw-snn8-svhn.csv};
					\draw[dashed, red, ultra thick] (axis cs:668,0) -- coordinate[pos=.7, pin={[pin distance=13pt, black, pin edge={solid, black}, outer sep=0pt, inner sep=0pt]25:\cnnname{8}}] (snnComp) (axis cs:668,28);
				\end{axis}
			\end{tikzpicture}
			\caption{FPS/W \snneightsvhn{} vs. \cnnname{8}}
		\end{subfigure}
		\caption{Comparison of energy and FPS/W between the \snnfoursvhn{} and \snneightsvhn{} and their corresponding CNN designs (\cnnname{7} and \cnnname{8}, respectively).}
		\label{fig:powersvhn}
	\end{figure}

	\begin{figure}
		\begin{subfigure}{.45\linewidth}
			\centering
			\begin{tikzpicture}
				\begin{axis}[
					ybar ,
					width=\linewidth,
					scaled x ticks=false,
					xmin=0.10,
					ymin=0,
					ylabel={\#input samples},
					xlabel={Power [W]}
					]
					\addplot+ [hist={bins=10}] table[y index=0] {data/hist_power-snn4-cifar.csv};
					\draw[dashed, red, ultra thick] (axis cs:0.587,0) -- coordinate[pos=.6, pin={[pin distance=19pt, black, pin edge={solid, thick, black}, outer sep=1pt, inner sep=0pt]150:\cnnname{9}}] (snnComp) (axis cs:0.587,40);
				\end{axis}
			\end{tikzpicture}
			\caption{Power \snnfourcifar{} vs. \cnnname{9}}
		\end{subfigure}
		\begin{subfigure}{.45\linewidth}
			\centering
			\begin{tikzpicture}
				\begin{axis}[
					ybar ,
					width=\linewidth,
					scaled x ticks=false,
					xmin=0.10,
					ymin=0,
					ylabel={\#input samples},
					xlabel={Power [W]}
					]
					\addplot+ [hist={bins=10}] table[y index=0] {data/hist_power-snn8-cifar.csv};
					\draw[dashed, red, ultra thick] (axis cs:0.687,0) -- coordinate[pos=.5, pin={[pin distance=17pt, black, pin edge={solid, thick, black}, outer sep=1pt, inner sep=0pt]150:\cnnname{10}}] (snnComp) (axis cs:0.687,50);
				\end{axis}
			\end{tikzpicture}
			\caption{Power \snneightcifar{} vs. \cnnname{10}}
		\end{subfigure}
		
		\vspace{1.5em}
		\begin{subfigure}{.45\linewidth}
			\centering
			\begin{tikzpicture}
				\begin{axis}[
					ybar ,
					width=\linewidth,
					scaled x ticks=false,
					xmin=0.00,
					ymin=0,
					xmax=8,
					ylabel={\#input samples},
					xlabel={Energy [mJ]}
					]
					\addplot+ [hist={bins=6}] table[y index=0] {data/hist_energy-snn4-cifar.csv};
					\draw[dashed, red, ultra thick] (axis cs:6.86,0) -- coordinate[pos=.65, pin={[pin distance=13pt, black, pin edge={solid, thick, black}, outer sep=0pt, inner sep=0pt]94:\cnnname{9}}] (snnComp) (axis cs:6.86,40);
				\end{axis}
			\end{tikzpicture}
			\caption{Energy \snnfourcifar{} vs. \cnnname{9}}
		\end{subfigure}
		\begin{subfigure}{.45\linewidth}
			\centering
			\begin{tikzpicture}
				\begin{axis}[
					ybar ,
					width=\linewidth,
					scaled x ticks=false,
					xmin=0.00,
					ymin=0,
					xmax=6,
					ylabel={\#input samples},
					xlabel={Energy [mJ]}
					]
					\addplot+ [hist={bins=6}] table[y index=0] {data/hist_energy-snn8-cifar.csv};
					\draw[dashed, red, ultra thick] (axis cs:5.08,0) -- coordinate[pos=.5, pin={[pin distance=13pt, black, pin edge={solid, thick, black}, outer sep=0pt, inner sep=0pt]95:{\cnnname{10}}}] (snnComp) (axis cs:5.08,60);
				\end{axis}
			\end{tikzpicture}
			\caption{Energy \snneightcifar{} vs. \cnnname{10}}
		\end{subfigure}
		
		\vspace{1.5em}    
		\begin{subfigure}{.45\linewidth}
			\centering
			\begin{tikzpicture}
				\begin{axis}[
					ybar ,
					width=\linewidth,
					scaled x ticks=false,
					xmin=0.04,
					ymin=0,
					ylabel={\#input samples},
					xlabel={FPS/W}
					]
					\addplot+ [hist={bins=10}] table[y index=0] {data/hist_fpsw-snn4-cifar.csv};
					\draw[dashed, red, ultra thick] (axis cs:145.6,0) -- coordinate[pos=.7, pin={[pin distance=13pt, black, pin edge={solid, thick, black}, outer sep=0pt, inner sep=0pt]95:\cnnname{9}}] (snnComp) (axis cs:145.6,40);
				\end{axis}
			\end{tikzpicture}
			\caption{FPS/W \snnfourcifar{} vs. \cnnname{9}}
		\end{subfigure}
		\begin{subfigure}{.45\linewidth}
			\centering
			\begin{tikzpicture}
				\begin{axis}[
					ybar ,
					width=\linewidth,
					scaled x ticks=false,
					xmin=0.04,
					ymin=0,
					ylabel={\#input samples},
					xlabel={FPS/W}
					]
					\addplot+ [hist={bins=10}] table[y index=0] {data/hist_fpsw-snn8-cifar.csv};
					\draw[dashed, red, ultra thick] (axis cs:196.7,0) -- coordinate[pos=.7, pin={[pin distance=13pt, black, pin edge={solid, black}, outer sep=0pt, inner sep=0pt]95:\cnnname{10}}] (snnComp) (axis cs:196.7,32);
				\end{axis}
			\end{tikzpicture}
			\caption{FPS/W \snneightcifar{} vs. \cnnname{10}}
		\end{subfigure}
		\caption{Comparison of energy and FPS/W between the \snnfourcifar{} and \snneightcifar{} and their corresponding CNN designs (\cnnname{9} and \cnnname{10}, respectively).}
		\label{fig:powercifar}
	\end{figure}

There is the possibility that not enough points in the value range are left for the encoding.
The condition for this is
\begin{equation}
2^{\left\lceil\log_2\frac{W}{K}\right\rceil} - \frac{W}{K} - 1 < 0.
\end{equation}
This occurs only when $\frac{W}{K}$ is approaching a power from two from below. In this rare case, we fall back to the original encoding.

Refer again to Table~\ref{tbl:improvements} for the effect of this \emph{compression} strategy.
Again a reduction of about 17\% of Watts can be observed.
Note that \snneightlutram{} and \snneightpacking{} show no difference because of the memory parallelism required here leads to already a minimum of BRAMs being used per PE in \snneightlutram{}.

Figure~\ref{fig:improvedpower} shows the resulting power estimations as well as total energy for one sample evaluation and the FPS/W for the MNIST dataset.
All metrics are again dependent on the input sample and therefore depicted as histograms.
As can be seen, the energy efficiency in terms of FPS/W is roughly similar for \snneightpacking{} and \snnfourpacking{}.
In the case of \snnfourpacking{}, energy consumption can be better in some cases but not in the average case than the comparable CNN implementation (\cnnname{5}).

\begin{table}
	\caption{Resource usage and vector-less power estimations of SNNs and CNNs for the SVHN dataset.}
    \resizebox{\columnwidth}{!}{%
\begin{tabular}{llNNNrrrrr}
	\toprule
  & & & & & \multicolumn{5}{c}{\tblheader{Power [W]}} \\ \cmidrule{6-10}
  \tblheader{Design} & \tblheader{Platform} & \tblheader{LUTs} & \tblheader{Regs.} & \tblheader{BRAMs} & \tblheader{Signals} & \tblheader{BRAMs} & \tblheader{Logic} & \tblheader{Clocks} & \tblheader{Total} \\ \midrule
  \cnnname{7} & PYNQ & 32765 &  50968 &  50 & 0.149 & 0.087 & 0.109 & 0.105 & \textbf{0.450} \\
  \cnnname{8} & PYNQ & 39927 &  59187 &  47.5 & 0.269 & 0.063 & 0.173 & 0.118 & \textbf{0.623} \\
  \cnnname{7} & \zcu & 32656 &  52964 &  46 & 0.225 & 0.053 & 0.263 & 0.202 & \textbf{0.743} \\
  \cnnname{8} & \zcu & 40172 &  59258 &  47 & 0.239 & 0.136 & 0.303 & 0.225 & \textbf{0.903} \\
  \snntwosvhn{} & PYNQ & 4733 & 2961 & 91 & 0.042 & 0.174 & 0.025 & 0.023 & \textbf{0.264} \\
  \snnfoursvhn{} & PYNQ & 9393 & 5652 & 92 & 0.068 & 0.175 & 0.043 & 0.036 & \textbf{0.322} \\
  \snneightsvhn{} & PYNQ & 18487 & 11024 & 104 & 0.146 & 0.200 & 0.091 & 0.063 & \textbf{0.500} \\
  \snnsixteensvhn{} & PYNQ & 37674 & 22077 & 140 & 0.348 & 0.265 & 0.185 & 0.116 & \textbf{0.914} \\
  \snntwosvhn{} & \zcu & 4896 & 2961 & 82 & 0.056 & 0.096 & 0.047 & 0.031 & \textbf{0.230} \\
  \snnfoursvhn{} & \zcu & 9293 & 5645 & 82 & 0.100 & 0.103 & 0.087 & 0.054 & \textbf{0.344} \\
  \snneightsvhn{} & \zcu & 18135 & 11013 & 100 & 0.204 & 0.163 & 0.181 & 0.104 & \textbf{0.652} \\
  \snnsixteensvhn{} & \zcu & 36038 & 21976 & 136 & 0.404 & 0.282 & 0.358 & 0.198 & \textbf{1.242} \\
  \bottomrule
\end{tabular}}%
\label{tbl:improvementssvhn}
\end{table}

\begin{table}
	\caption{Resource usage and vector-less power estimations of SNNs and CNNs for the CIFAR-10 dataset.}
	\resizebox{\columnwidth}{!}{%
		\begin{tabular}{llNNNrrrrr}
			\toprule
			& & & & & \multicolumn{5}{c}{\tblheader{Power [W]}} \\ \cmidrule{6-10}
			\tblheader{Design} & \tblheader{Platform} & \tblheader{LUTs} & \tblheader{Regs.} & \tblheader{BRAMs} & \tblheader{Signals} & \tblheader{BRAMs} & \tblheader{Logic} & \tblheader{Clocks} & \tblheader{Total} \\ \midrule
			\cnnname{9} & PYNQ & 30745 &  42436 &  73 & 0.279 & 0.084 & 0.125 & 0.99 & \textbf{0.587} \\
			\cnnname{10} & PYNQ & 38111 &  64962 &  75.5 & 0.309 & 0.089 & 0.175 & 0.114 & \textbf{0.687} \\
            \cnnname{9} & \zcu & 30848 &  43075 &  48 & 0.282 & 0.088 & 0.289 & 0.231 & \textbf{0.890} \\
            \cnnname{10} & \zcu & 38447 &  66797 &  50 & 0.292 & 0.092 & 0.343 & 0.243 & \textbf{0.970} \\
			\snntwocifar{} & PYNQ & 2566 & 25151 & 118 & 0.115 & 0.217 & 0.056 & 0.050 & \textbf{0.438} \\
			\snnfourcifar{} & PYNQ & 5063 & 27504 & 136 & 0.122 & 0.313 & 0.076 & 0.052 & \textbf{0.563} \\
			\snneightcifar{} & PYNQ & 21245 & 44126 & 140 & 0.179 & 0.321 & 0.103 & 0.061 & \textbf{0.664} \\
			\snntwocifar{} & \zcu & 4925 & 2962 & 146 & 0.057 & 0.135 & 0.046 & 0.036 & \textbf{0.274} \\
			\snnfourcifar{} & \zcu & 9595 & 5655 & 146 & 0.103 & 0.142 & 0.088 & 0.058 & \textbf{0.391} \\
			\snneightcifar{} & \zcu & 18199 & 11016 & 164 & 0.203 & 0.202 & 0.181 & 0.109 & \textbf{0.695} \\
			\snnsixteencifar{} & \zcu & 36115 & 21982 & 200 & 0.399 & 0.320 & 0.356 & 0.205 & \textbf{1.280} \\
			\bottomrule
	\end{tabular}}%
	\label{tbl:improvementscifar}
\end{table}

Figures~\ref{fig:powersvhn} and~\ref{fig:powercifar} present the same charts of results for the SVHN and CIFAR-10 datasets, respectively.
SNNs are named after their parallelization factor and the dataset whose corresponding model they implement (see Table~\ref{tbl:nn-models}).
Resource usage and corresponding power results received using vector-less estimation are shown in Tables~\ref{tbl:improvementssvhn} and~\ref{tbl:improvementscifar}.
All results were synthesized for the PYNQ board with a clock frequency of 100~MHz, while for the ZCU102, we consistently used a frequency of 200~MHz.
Note that this affects the power estimations, which are according to the corresponding frequency setting.

The NN architecture used for SVHN has more than 14 times as many weights as well as the need for larger membrane potential memories compared to the network for the MNIST data.
This is why both power and latency measurements are higher than in the MNIST case.
The same holds true for CIFAR-10.
The CNN designs considered have been chosen to have almost equal estimated power values as the SNNs.
Similar to the MNIST dataset, CNNs use more registers and fewer BRAMs for storing intermediate values between layers.
However, this leads to corresponding decreases and increases in the BRAM and Signal categories of the estimated power.
Likewise, CNNs use more LUTs because they are instantiated as part of MAC units, while for SNNs, LUTs are employed predominantly as memory which is also restricted, e.g., 17,400 LUT slices being available on the xc7Z020 FPGA.

Also, one major difference between FINN generated CNN implementations and the synthesized SNN implementations is that FINN uses a dedicated streaming dataflow architecture.
This means that an IP block is instantiated on the FPGA for each layer.
The more layers there are in a network, the fewer options remain for configuring and optimizing the throughput of bottleneck parts of the network.
This can be seen when looking at the latencies needed to process one input sample shown in Figure~\ref{fig:latencysvhn}.
When comparing CNN and SNN implementations having approximately equal power estimations, the CNN equivalents \cnnname{7} and \cnnname{8} as well as \cnnname{9} and \cnnname{10} become slower in comparison.

	\begin{figure}
		\begin{subfigure}{.49\linewidth}
			\begin{tikzpicture}
				\begin{axis}[
					ybar,
					width=\linewidth,
					xmin=0,
					ymin=0,
					ylabel={\#input samples},
					scaled x ticks = false,
					xticklabel style={rotate=45, anchor=east},
					xlabel={Latency $\times$100}
					]
					\addplot+ [hist={bins=10}] table[y index=0] {data/hist_latency-snn4-svhn.csv};
					\draw[dashed, red, ultra thick] (axis cs:6248,0) -- 
					coordinate[pos=.7, pin={[black,pin distance=15pt, pin edge={solid, thick, black}, outer sep=0pt, inner sep=0pt]45:\cnnname{7}}] (snnComp) 
					(axis cs:6248,330);
				\end{axis}
			\end{tikzpicture}
			\caption{\snnfoursvhn{} vs.\ \cnnname{7}}
		\end{subfigure}
		\begin{subfigure}{.49\linewidth}
			\begin{tikzpicture}
				\begin{axis}[
					ybar ,
					width=\linewidth,
					xmin=0,
					ymin=0,
					ylabel={\#input samples},
					xlabel={Latency $\times$100},
					xticklabel style={rotate=45, anchor=east}
					]
					\addplot+ [hist={bins=10}] table[y index=0] {data/hist_latency-snn8-svhn.csv};
					\draw[dashed, red, ultra thick] (axis cs:3829,0) -- 
					coordinate[pos=.7, pin={[black,pin distance=9pt,pin edge={solid,thick, black}, outer sep=0pt, inner sep=0pt]10:\cnnname{8}}] (snnComp) 
					(axis cs:3829,330);
				\end{axis}
			\end{tikzpicture}
			\caption{\snneightsvhn{} vs. \cnnname{8}}
		\end{subfigure}
		\begin{subfigure}{.49\linewidth}
			\begin{tikzpicture}
				\begin{axis}[
					ybar,
					width=\linewidth,
					xmin=0,
					ymin=0,
					scaled x ticks = false,
					ylabel={\#input samples},
					xlabel={Latency $\times$100},
					xticklabel style={rotate=45, anchor=east}
					]
					\addplot+ [hist={bins=10}] table[y index=0] {data/hist_latency-snn4-cifar.csv};
					\draw[dashed, red, ultra thick] (axis cs:11715,0) -- 
					coordinate[pos=.8, pin={[black,pin distance=40pt, pin edge={solid, thick, black}, outer sep=0pt, inner sep=0pt]180:\cnnname{9}}] (snnComp) 
					(axis cs:11715,480);
				\end{axis}
			\end{tikzpicture}
			\caption{\snnfourcifar{} vs.\ \cnnname{9}}
		\end{subfigure}
		\begin{subfigure}{.49\linewidth}
			\begin{tikzpicture}
				\begin{axis}[
					ybar ,
					width=\linewidth,
					xmin=0,
					ymin=0,
					xmax=8000,
					ylabel={\#input samples},
					xlabel={Latency $\times$100},
					xticklabel style={rotate=45, anchor=east}
					]
					\addplot+ [hist={bins=10}] table[y index=0] {data/hist_latency-snn8-cifar.csv};
					\draw[dashed, red, ultra thick] (axis cs:7411,0) -- 
					coordinate[pos=.6, pin={[black,pin distance=6pt,pin edge={solid, thick, black}, outer sep=0pt, inner sep=0pt]95:\cnnname{10}}] (snnComp) 
					(axis cs:7411,630);
				\end{axis}
			\end{tikzpicture}
			\caption{\snneightcifar{} vs. \cnnname{10}}
		\end{subfigure}
		\caption{Latency comparison of SNN implementations performing SVHN and CIFAR-10 classification with parallelization factor $P=4$ and $P=8$ each. The histograms were generated by measuring the latency for 1,000 images taken from each dataset.}
		\label{fig:latencysvhn}
\end{figure}
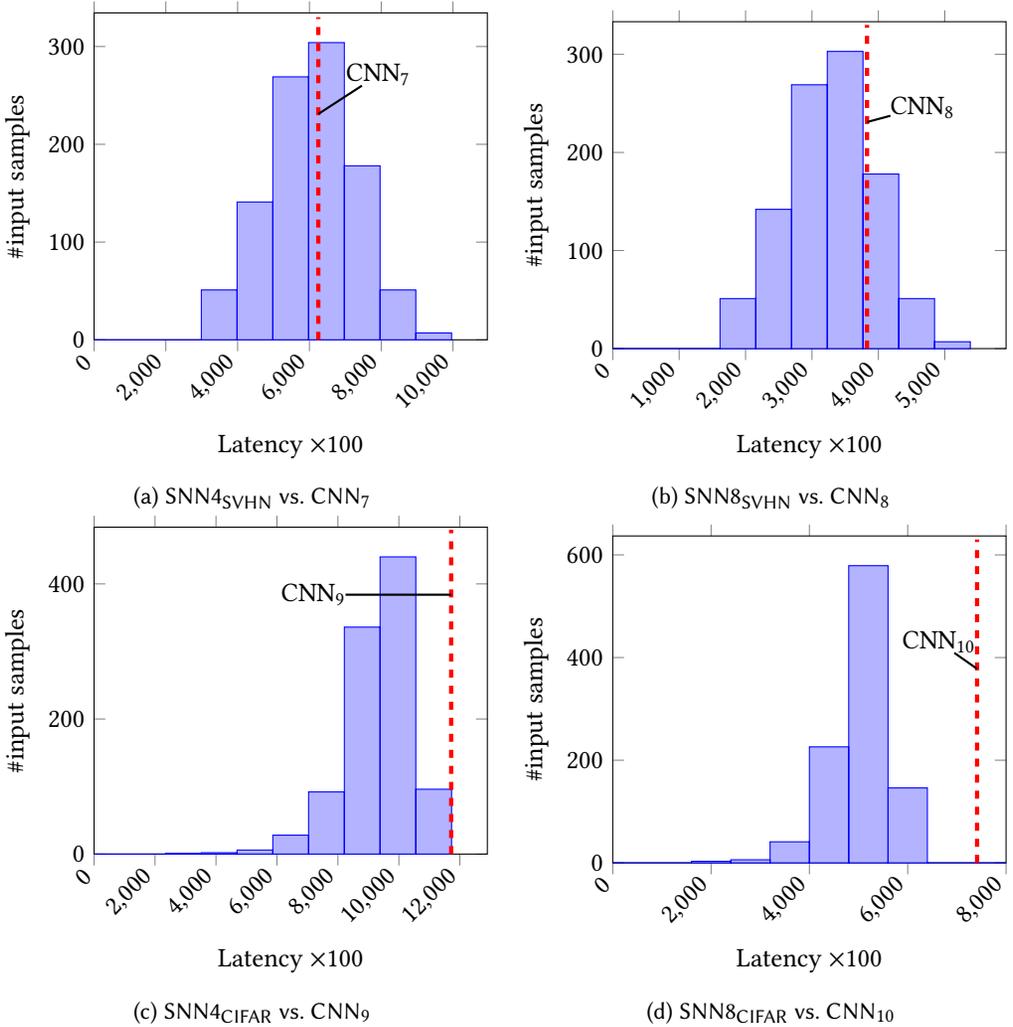

For more than half of the input samples, \snneightsvhn{} needs less energy than \cnnname{8}.
For the larger network model (CIFAR-10), \snneightcifar{} has a higher energy efficiency than \cnnname{10}.

Moreover, for both, MNIST and CIFAR-10, SNNs with \(P=8\) yield the best energy efficiency.
Since the ZCU102 board has a different chip technology and architecture than the PYNQ board, BRAMs use less power in this case.
However, clock routing is more expensive in terms of energy compared to the PYNQ platform.
With increasing parallelization factor $P$, the ZCU102 scales a little worse than the PYNQ.
For example, \snnsixteensvhn{} consumes more power due to the Clocks category on the ZCU102 than on the PYNQ.

On the other hand, memory resources soom become too scarce on the PYNQ for \snneightcifar{}, so registers must be used.
Likewise, \snnsixteencifar{} cannot be implemented on the PYNQ board due to the resource requirements.

The FINN-based CNN implementations witness an increased dynamic power on the larger ZCU102 board when compared to the PYNQ platform.
This is due to the use of LUTs and Registers for MAC operations, whereas in the SNN accelerator, they are used to a much larger degree for keeping intermediate results or storing read-only weights.

Table~\ref{tbl:relatedquant} compares existing SNN implementations with our implementations in terms of classification accuracy and FPS/W for the MNIST, SVHN, and CIFAR-10 datasets.
Works discussed in the related work Section~\ref{sec:rel_work} but not listed in Table~\ref{tbl:relatedquant} either did not provide FPS/W data or reported results for networks/data sets not considered in this work.

\begin{table}
	\caption{Overview of related SNN accelerator approaches for multiple data sets with respect to accuracy and FPS/W. Empty cells in the related work indicate that these values are not available. The accelerators \snnfourlutram{} and \snneightlutram{} have not been used for the SVHN and CIGAR-10 benchmarks as they are not optimized. For the CIFAR-10 data set, the \snnsixteenpacking{} has a resource requirement that the Pynq board cannot meet and, hence, no results are reported.}
	\label{tbl:relatedquant}
    \resizebox{\columnwidth}{!}{%
		\pgfplotstabletypeset[
		columns/Work/.style={string type,column type={l}},
		columns/Platform/.style={string type,column type={c}},
		columns/Spike Encoding/.style={string type},
		columns/FPS/W MNIST/.style={
			string type,
			column type={r},
			column name={FPS/W},
			assign cell content/.append code={
				\ifnum\pgfplotstablerow<11
				\pgfkeyssetvalue{/pgfplots/table/@cell content}{##1\phantom{]}}%
				\else
				\pgfkeyssetvalue{/pgfplots/table/@cell content}{##1}%
				\fi
			},
		},
		columns/FPS/W SVHN/.style={
			string type,
			column type={r},
			column name={FPS/W},
			empty cells with={--},
			assign cell content/.append code={
				\ifnum\pgfplotstablerow<11
				\pgfkeyssetvalue{/pgfplots/table/@cell content}{\ifstrequal{##1}{}{--}{##1\phantom{]}}}%
				\else
				\pgfkeyssetvalue{/pgfplots/table/@cell content}{##1}%
				\fi
			},
		},
		columns/FPS/W CIFAR/.style={
			string type,
			column type={r},
			column name={FPS/W},
			empty cells with={--},
			assign cell content/.append code={
				\ifnum\pgfplotstablerow<11
				\pgfkeyssetvalue{/pgfplots/table/@cell content}{\ifstrequal{##1}{}{--}{##1\phantom{]}}}%
				\else
				\pgfkeyssetvalue{/pgfplots/table/@cell content}{##1}%
				\fi
			},
		},
		every head row/.style={
			before row={%
				\toprule
				& & \multicolumn{2}{c}{\textbf{MNIST}} & \multicolumn{2}{c}{\textbf{SVHN}} & \multicolumn{2}{c}{\textbf{CIFAR-10}}\\
				\cmidrule{3-8}
			},
			after row=\midrule,
		},
		columns={Work, Platform, AccuracyMNIST, FPS/W MNIST, AccuracySVHN, FPS/W SVHN, AccuracyCIFAR, FPS/W CIFAR},
		every row no 1/.style={after row=\midrule},
		every row no 7/.style={after row=\midrule},
		skip rows between index={0}{1},
		skip rows between index={2}{3},
		skip rows between index={6}{7},
		skip rows between index={11}{12},
		]{\relWorkTable}}
\end{table}

To generate data for SyncNN~\cite{syncnn}, we use the open-source code\footnote{\url{https://github.com/SFU-HiAccel/SyncNN}} provided by the authors, scaled down the LeNet-S configuration and synthesized it for the PYNQ-Z1.
This uses 16,326~LUTs and 16,228 registers along with 69~DSPs and 253~half~BRAMs.
For comparability, we then use the vector-less Vivado Power Estimator tool to measure a dynamic power of 0.405 W.
Together with the reported 800 FPS for the ZedBoard~\cite{syncnn}, this yields an energy efficiency of 1975 FPS/W on the PYNQ-Z1. 
We likewise read the throughput for the same network architecture applied to the SVHN dataset as 90 FPS, arriving at 222 FPS/W.
For CIFAR-10, we synthesize SyncNN with an 8-bit NiN network \cite{nin} configuration, which is estimated to consume 0.553 W.

The values for the other SNNs have been taken from the respective publication.

As can be seen, together with the applied improvements, the examined architecture is a state-of-the-art accelerator for \acp{SNN} on embedded platforms.
Regarding SVHN and CIFAR-10, only FireFly achieves an energy efficiency which falls into the intervals measured for \snneightcifar{}.

From Table~\ref{tbl:relatedquant}, we can recognize a lower classification rate for the SNN implementations for the larger networks when using the \texttt{snntoolbox}. In the future, we would like to investigate alternative ways for SNN training such as done by Cerebron~\cite{cerebron} with which we hope to obtain similarly high accuracies.

\section{Conclusion}\label{sec:conclusion}
In this work, we analyzed whether \acp{SNN} really offer a promised higher energy efficiency in comparison to conventional \acp{CNN} as they are sometimes marketed. For this, comparisons between different \ac{CNN} and \ac{SNN} implementations have been carried out to find an confirmation of this hypothesis when targeting FPGA devices.
It can be shown that \acp{SNN} can be faster in some cases but can fall short in average power consumption for smaller classification tasks such as MNIST. 


As candidates, we compared CNN architectures synthesized using the FINN-based streaming dataflow architecture with a parameterizable SNN architecture introduced in~\cite{sommer} for two FPGA platforms of different size and the three benchmark data sets MNIST, SVHN and CIFAR-10. 

We also investigated potential techniques to reduce the power footprint of the SNN architecture.
This was done, first, by instantiating LUTRAM instead of BRAMs to store address events and, second, by employing an improved encoding scheme for spike events.
These ideas have led to a total increase in energy efficiency (FPS/W) by a factor of 1.41 for the MNIST case.

For the comparison of different pairs of CNN and SNN nets for a given benchmark and FPGA platform, we matched solutions of equal power. To finally answer our initial question of whether to spike or not to spike, we showed that for small scale benchmarks such as MNIST, matching SNN designs provide rather no or little energy efficiency improvements. For large networks such as used for the SVHN and CIFAR-10 data sets, the trend reverses.
The reason for this is that MAC units as well as FIFO buffers instantiated for each layer for CNN implementations synthesized using the FINN-based streaming dataflow architecture principle, incur a high power consumption such that the SNN implementations provide a higher average FPS/W.



\begin{acks}
This work was supported by the Deutsche Forschungsgemeinschaft (DFG, German Research Foundation) -- 450987171.
\end{acks}

%
%
\printbibliography

\end{document}